\documentclass{article}

\usepackage{style}
\usepackage{formulas_style}

\title{Barbero--Immirzi--Holst Lagrangian with Spacetime Barbero--Immirzi Connections}
\date{}
\author{
Andrea Orizzonte
	\\
{\small Dipartimento di Matematica, University of Torino, via Carlo Alberto 10, 10123 Torino (Italy)
	\\
e-mail: \href{mailto:andrea.orizzonte@unito.it}{andrea.orizzonte@unito.it} }
}

\begin{document}

%%% intro
%\frontmatter
%\include{./tex/fronte_lm}
%
%\include{./tex/0_intro}
%
%\setcounter{tocdepth}{1}
%\tableofcontents

%%% main
%\mainmatter

\maketitle

\begin{abstract}

We carry out the complete variational analysis of the Barbero--Immirzi--Holst Lagrangian, which is the Holst Lagrangian expressed in terms of the triad of fields $(\theta, A, \kappa)$, where $\theta$ is the solder form/spin frame, $A$ is the spacetime Barbero--Immirzi connection, and $\kappa$ is the extrinsic spacetime field. The Holst Lagrangian depends on the choice of a real, non zero Holst parameter $\gamma \neq 0$ and constitutes the classical field theory which is then quantized in Loop Quantum Gravity. The choice of a real Immirzi parameter $\beta$ sets up a one-to-one correspondence between pairs $(A, \kappa)$ and spin connections $\omega$ on spacetime. The variation of the Barbero--Immirzi--Holst Lagrangian is computed for an arbitrary pair of parameters $(\beta, \gamma)$. We develop and use the calculus of vector-valued differential forms to improve on the results already present in literature by better clarifying the geometric character of the resulting Euler--Lagrange equations. The main result is that the equations for $\theta$ are equivalent to the vacuum Einstein Field Equations, while the equations for $A$ and $\kappa$ give the same constraint equation for any $\beta \in \R$, namely that $A + \kappa$ must be the Levi--Civita connection induced by $\theta$. We also prove that these results are valid for any value of $\gamma \neq 0$, meaning that the choice of parameters $(\beta, \gamma)$ has no impact on the classical theory in a vacuum and, in particular, there is no need to set $\beta = \gamma$.

%\keywords{
\eop \eop \noindent \textbf{Keywords:}
Vector-valued differential forms, principal connections, structure group reduction, variational calculus, Loop Quantum Gravity
%}
%\and More}

\end{abstract}

\section{Introduction and Results}

The Holst Lagrangian is dependent on a real, non zero parameter $\gamma$ called the \emph{Holst parameter} and it is used as the starting, classical theory of gravitation to be quantized in Loop Quantum Gravity (LQG), see, for example, \cite{rovellibook,rovelli_vidotto}. The Holst lagrangian is defined on a connected, orientable, $4$-dimensional lorentzian manifold $M$ and it depends on a spin frame/solder form $\theta^a$ at order zero and on a spin connection $\omega^{ab}$ at first order, explicitly it is (see \cite{ffr2})
\eq{
\sv L_\gamma(\theta, j^1 \omega) = \frac{1}{4\bar G} \[ \epsilon_{abcd} R^{ab} \w \theta^c \w \theta^d + \frac{2}{\gamma} R^{ab}{} \w \theta^\._{a} \w \theta^\._{b} \], \quad \gamma \neq 0
}
where $R^{ab}$ is the Riemann tensor of $\omega^{ab}$, $\bar G$ includes all physical constants, and latin indices are raised/lowered via the Minkwoski metric $\eta$ on $\R^4$. Notice that the first term in $\sv L_\gamma$ is the Hilbert--Einstein Lagrangian in the frame-affine formalism (i.e.\ where the spin frame and spin connections are treated as two separate, independent variables).

The original definition was given by Holst \cite{holst_lagr}, the motivation was to have a Lagrangian which is classically equivalent to the Hilbert--Einstein Lagrangian but for which thet pair of canonical variables is $(A^k_A(\gamma), E^A_k)$, where $A^k_A(\gamma)$ is the Barbero--Immirzi (BI) connection with parameter $\gamma$ and $E^A_k$ is the densitized triad. The BI connection was defined by Barbero \cite{barbero94} and Immirzi \cite{immirziqgrc} in order to deal with the canonical constraints of general relativity.

We now briefly recall this setting. Fix a $4$-dimensional orientable lorentzian manifold $M$ with an Einstein metric $g$ and an embedded $3$-dimensional submanifold $S$ that is spacelike, so that the pull-back metric $h = t^* g$ along the embedding map $t \colon S \hookrightarrow M$ is positive definite.
Choose local coordinates $\{ x^\mu \}_{\mu = 0, \dots, 3}$ on $M$, local coordinates $\left\{ s^A \right\}_{A = 1,2,3}$ on $S$, and denote by $\{ T_a \}_{a = 0, \dots, 3}$ the standard basis in $\R^4$ which is orthonormal with respect to the Minkowski metric $\eta$, that is
\eq{
\eta(T_a, T_b) = \eta_{ab} = \sys{
-1 & \text{if } a = b = 0
	\\
1 & \text{if } a = b = 1,2,3
	\\
0 & \text{if } a \neq b 
}
}
The metric $g$ can be alternatively described in terms of \emph{tetrads $e^a_\mu$} with
\eq{
g_{\mu\nu} = e^a_\mu \, \eta_{ab} \, e^b_\nu
}
By pullback throught $t$ we can define \emph{triads $e^i_A$} on $S$, with $i = 1,2,3$, and the pull-backed metric $h$ can the be expressed as
\eq{
h_{AB} = e^i_A \, \delta_{ij} \, e^j_B
}
The inverse matrices of $e^a_\mu$ and $e^i_A$ are $\theta^\mu_a$ and $\theta^A_i$.
%and we define the \emph{densitized tetrad $E^\mu_A$ and triad $E^A_i$} as
%\eq{
%E^\mu_a &= \sqrt{\abs{\det g}} e^\mu_a
%	\\
%E^A_i &= \sqrt{\abs{\det h}} e^A_i
%}
From the canonical analysis of the Hilbert--Einstein Lagrangian we get (see, for example,\cite{barbero94}) that the canonical variables are
\eq{
\sys{
\tilde \Gamma^k_A = \half 1 \epsilon_{ij}{}^k \, \Gamma^{ij}_A
	\\
E^A_i = \sqrt{\abs{\det h}} \, \theta^A_i
}
}
where $\Gamma^{ij}_A$ are the Christoffel symbols for $h$ and $E^A_i$ is called densitized triad. The Holst Lagrangian with Holst parameter $\gamma$ adds a term to the Hilbert--Einstein Lagrangian so that the resulting canonical variables are
\eq{
\label{bi_conn_defn}
\sys{
A^k_A(\gamma) = \tilde \Gamma^k_A + \gamma \kappa^k_A = \half 1 \epsilon_{ij}{}^k \, \Gamma^{ij}_A + \gamma \kappa^{k}_A
	\\
E^A_i = \sqrt{\abs{\det h}} \, \theta^A_i
}
}
The $\su(2)$-connection $A^k_A$ on $S$ is called \emph{Barbero--Immirzi (BI) connection with parameter $\gamma$}, and it is defined out of the Levi--Civita connection $\Gamma^{ij}_A$ of $h$ and the $\kappa^k_A$, which are related to the coefficients of the Weingarten operator of $S$ (see \cite{kn2},p.\ 14). The BI connection coefficients $A^k_A$ are the lorentzian equivalent of the Ashtekar variables, which where introduced by Ashtekar in \cite{ashtekar_var} in order to treat the canonical analysis of constraints in the quantization of euclidean gravity.

The use of BI connections $A^k_A$ as canonical variables is due to that analysis of constraints retain some good p

As we mentioned the Euler--Lagrange (E--L) equations for the Holst Lagrangian are equivalent to that of the Hilbert--Einstein Lagrangian, so that they describe the same \emph{classical} theory, but the presence of a non zero Holst parameter $\gamma$ has a heavy influence on the resulting quantum theory and on the coupling with matter, see \cite{rovellibook,rovelli_vidotto} for a detailed account.

The issue with canonical analysis and canonical quantization is that by treating time and space separately it breaks one important feature of relativistic theories, namely general covariance, so that a different approach is required to find a fully covariant quantization of general relativity and of relativistic theories in general.

One of the first problems to be solved in this direction, is that of performing a complete, covariant analysis of the Holst Lagrangian as a \emph{classical} theory, which is then to be used as a solid foundation for a covariant quantization scheme. This is done in two steps. First, we study the \emph{kinematic} aspects of the theory, that is, we give a geometric/covariant characterization of the fields on which the Lagrangian depends. Second, we carry out the covariant variational analysis of the Holst Lagrangian, that is we study the classical \emph{dynamics}.

The main difficulty in the kinematical analysis is to define BI connections \emph{on spacetime}, since in \cref{bi_conn_defn} the $\su(2)$-connection is only defined on the space submanifold $S$. We already dealt with this in the previous paper \cite{orizz_bi_conn}, in which we built and characterized BI connections on spacetime as fully covariant objects. The main result of this work is that one can always define a spacetime BI connection out of a given spin connection in any $(n+1)$-dimensional lorentzian manifold $M$ by choosing a reductive splitting for the pair of Lie algebras $\( \spin(n,1), \spin(n) \)$. The spacetime BI connection will then be the $\spin(n)$-component of the spin connection $\omega$. For $n = 3$ one gets that there is a one-parameter family of reductive splittings
\eq{
\spin(3,1) = \su(2) \oplus \lie m_\beta, \quad \beta \in \R
}
If $\omega = A(\beta) \oplus \kappa(\beta)$ in the splitting above, we have the local expressions
\eq{
\sys{
A^k(\beta) = \half 1 \epsilon_{ij}{}^k \, \omega^{ij} + \beta \, \omega^{0k}
	\\
\kappa^k(\beta) = \omega^{0k}
}
}
$A$ is the spacetime BI connection, $\kappa$ is the \emph{exterior spacetime field}, and the real splitting parameter $\beta$ is called \emph{Immirzi parameter}. Even though the expression above is analoguous to \cref{bi_conn_defn}, the Immirzi parameter $\beta$ is kinematical, since it depends only on the geometric character of BI connections, while the Holst parameter $\gamma$ is dynamical since it depends on the choice of the Holst Lagrangian to describe gravity. 

The two parameters $\beta$ and $\gamma$ seem, from this point of view, completely unrelated. Different possibilites arise. The first is that the relation between $\beta$ and $\gamma$ may be dictated by the form of the E--L equations of the Holst Lagrangian, that is, the equations may select one particular reductive splitting $\spin(3,1) = \su(2) \oplus \lie m_\beta$ by constraining the dependency of $\beta$ on $\gamma$. The second possibility is that both $\beta$ and $\gamma$ carry a physical significance, in coupling different degrees of freedom in the vacuum theory or as coupling constants when the Holst Lagrangian is combined with matter lagrangians. 

When we express the Holst Lagrangian in terms of $(\theta, A, \kappa)$, instead of $(\theta, \omega)$, we speak of the \emph{Barbero--Immirzi--Holst (BIH) Lagrangian}. We clearly expect the BIH Lagrangian to be equivalent to the Holst Lagrangian, since a pair $(A, \kappa)$ can be used to reconstruct a spin connection $\omega = A + \kappa$. Also, we expect equivalence to the Hilbert--Einstein Lagrangian, by transitivity. This was in part already verified in a work by Fatibene, Francaviglia, and Rovelli in \cite{ffr2}. This is done by splitting the pair of spin indices $ab$ ($a,b = 0, \dot, 3$) into their $\su(2)$ indices $ij$ ($i,j = 1, 2, 3$) and their $\lie m_\beta$ pairs $0k$ ($k = 1,2,3$), the resulting E--L equations can be shown to be equivalent to the Einstein field equations, but their geometrical significance is far from clear and, moreover, it is assumed $\beta = \gamma$ from the beginning.

The aim of the present work then is threefold: to carry out the covariant variational analysis of the Holst Lagrangian with the triple $(\theta, A, \kappa)$ as fundamental fields, to extract geometric information from the resulting E--L equations, and to investigate the role and significance of the parameters $\beta, \gamma$ in the resulting vacuum field equations. To this end we develop and use the calculus of valued-valued forms, which aid both in streamlining the calculations involved in the variational analysis and in the investigation of the geometrical meaning of the objects we encounter. We succeed in proving that, in the vacuum theory, the resulting E--L equations pose no constraint on the values of $\beta$ and $\gamma$, on the other hand two of the three sets of equations give strong constraints on the relation between the BI connection $A$ and the extrinsic spacetime field $\kappa$. 
%The variational analysis is also very straightforward and it very adaptable to the case of matter coupling, especially spinors.

In \cref{prereq} we list the most important definitions and results of \cite{orizz_bi_conn}, along with some results on spin connections. These are here mainly as a quick reference for the rest of the work.

In \cref{vec_forms_vec_forms} we give the definition and basic properties of vector-valued forms on a spin bundle $\sv Q = (Q, \pi, M, \Spin_0(r,s))$ over the spacetime manifold $M$. 

In \cref{vec_forms_hodge} we specialize to the case of tensorial forms and the definition of a suitable generalization of Hodge duality on these forms: the composite Hodge dual.

In \cref{vec_forms_trace_sec} we use the newly defined composite Hodge dual to give an intrinsic notion of trace on tensorial forms. We also study the relation between the trace and the other operations on vector-valued forms.

The results of the preceeding three sections are essentially original. Some concepts where already present in a similar form in the book by Besse \cite{besse_french}, albeit for the special case of the tangent manifold $TM$. We have tried to keep the same nomenclature for continuity and to make the generalization more explicit.

In \cref{bi_dynamical_holst} we begin our study of the dynamics by recasting the Holst Lagrangian in the language of vector-valued forms. We introduce the operator $\st_\gamma$ and characterize it.

In \cref{bi_dynamical_bih} we pass to the Barbero--Immirzi--Holst Lagrangian by changing the fundamental fields from $(\theta, \omega)$ to $(\theta, A, \kappa)$.

Finally, in \cref{bi_dynamical_bih_el}, we carry out the variational analysis of the Barbero--Immirzi--Holst Lagrangian for any pair of parameters $(\beta, \gamma)$. We write the resulting Euler--Lagrange equations explicitly and discuss their geometrical implications. We show the dynamical equivalence with Einstein Field Equations and show that classically the parameters $\beta$ and $\gamma$ have no relevance in the vacuum case.

In the \cref{app_trace_lemma} we give the proof of the trace lemma and the statement regarding the injectivity of the trace, which are at the end of \cref{vec_forms_trace_sec}. 

\newpage
\section{Spin Frames, Spin Connections, and Barbero--Immirzi Connections}\label[sec]{prereq}

\subsection{Spin Group, Spin Algebra, and Reductive Pairs}

Some of the material contained in this subsection can be found in \cite{orizz_bi_conn} and references therein.

Denote by $\eta$ the standard metric in $\R^m$ of signature $(r,s)$ and by $\{ T_a \}_{a = 1, \dots, m}$ the standard $\eta$-orthonormal basis of $\R^m$, that is
\eq{
\eta(T_a, T_b) = \eta_{ab} = \sys{
-1 & \text{if } a = b = 1, \dots, s
	\\
1 & \text{if } a = b = s+1, \dots, s+r = m
	\\
0 & \text{if } a \neq b
}
}

The \emph{orthogonal group (in signature $(r,s)$)} is denoted by $\O(r,s)$ and is
\eq{
\O(r,s) 
&= \{ \Lambda \in \GL(m) : \eta(\Lambda v , \Lambda w) = \eta(v, w), \quad \forall v, w \in \R^m \}
	\\
&= \{ \Lambda \in \GL(m) : \T\Lambda \, \eta \, \Lambda = \eta \}
}
The special orthogonal group $\SO(r,s)$ is the subgroup
\eq{
\SO(r,s) = \{ \Lambda \in \O(r,s) : \det \Lambda = 1 \}
}

It is known that for euclidean signatures $(r,s) = (m,0)$ the orthogonal group has two connected components, with $\SO(m) = \SO(m,0)$ being the connected component of the identity. In indefinite signatures $(r,s)$ and dimension $m \geq 3$, however, the orthogonal group has four connected components. The connected component of the identity is denoted by $\SO_0(r,s)$ and it is a strict subgroup of the special orthogonal group, that is $\SO_0(r,s) \subsetneq \SO(r,s)$.

Denote by $\Cl(r,s)$ the Clifford algebra of $\R^m$ with respect to the quadratic form induced by $\eta$, as is customary we denote by boldface $\mb v$ the image of a vector $v \in \R^m$ in the Clifford algebra. The \emph{spin group (of signature $(r,s)$)} is
\eq{
\Spin(r,s) = \{ S \in \Cl(r,s) : S = \mb v_1 \cdots \mb v_{2k}, v_i \in \R^m, \eta(v_i,v_i) = \pm 1 \}
}
That is, elements in $\Spin(r,s)$ are products of an even number of unitary vectors in the Clifford algebra. For any unitary vector $v \in \R^m$ denote the reflection through $v$ as $\ell(v)$, the corresponding map in the Clifford algebra is
\eq{
\ell(v) \mb w = -\eta(v,v) \, \mb v \mb w \mb v
}
This defines a map from $\Spin(r,s)$ to $\SO(r,s)$ which is
\eq{
\ell(S = \mb v_1 \cdots \mb v_{2k}) = \ell(\mb v_1) \circ \cdots \circ \ell(\mb v_{2k})
}
The map $\ell \colon \Spin(r,s) \rarr \SO(r,s)$ can be proven to be a two-to-one covering map. In the euclidean ($(r,s) = (m,0)$) and lorentzian ($(r,s) = (m-1, 1)$) signatures the spin group is also the universal covering group of the special orthogonal group. By the properties of covering maps, the preimage of $\SO_0(r,s)$ through $\ell$ is also the connected component of the identity in $\Spin(r,s)$, denote it by $\Spin_0(r,s)$.

Since $\SO(r,s)$ and $\Spin(r,s)$ are Lie groups their Lie algebras $\so(r,s)$ and $\spin(r,s)$ are isomorphic Lie algebras. Recall that we have the vector space isomorphism
\eq{
\so(r,s) \simeq \Lambda^2 \R^m
}
where the bivector $x \w y \in \Lambda^2 \R^m$ acts as
\eq{
(x \w y)(v) = \eta(x,v) \, v - \eta(y,v) \, x, \quad \forall v \in \R^m
}

Since
\eq{
\spin(r,s) = \{ X \in \Cl(r,s) : X = \mb x \mb y, x, y \in \R^m \}
}
we get the vector space isomorphism $\Psi \colon \spin(r,s) \rarr \Lambda^2 \R^m$ given by
\eq{
&\Psi(\mb x \mb y) = -2 \, x \w y, \quad \text{for }\eta(x, y) = 0
	\\
&\inv \Psi(x \w y) = -\frac{1}{4} [\mb x, \mb y]
}

Using the $\Psi$ isomorphism we can induce a Lie bracket on $\Lambda^2 \R^m$, giving it the structure of Lie algebra and turning $\Psi$ into a Lie algebra isomorphism. We can also induce a scalar product on $\spin(r,s)$ from the canonical one on $\Lambda^2 \R^m$. Denote this scalar product by $q$, then this is given on the orthonormal basis as
\eq{
q(\mb T_a \mb T_b, \mb T_c \mb T_d) 
&= \eta(-2 \, T_a \w T_b, -2 \, T_c \w T_d)
	\\
&= 4 \begin{vmatrix} \eta(T_a, T_c) & \eta(T_a, T_d) \\ \eta(T_b, T_c) & \eta(T_b, T_d) \end{vmatrix}
	\\
&= 4 \( \eta_{ac} \eta_{bd} - \eta_{ad} \eta_{bc} \)
}
We define the $q$-orthonormal basis $\{ J_{ab} \}$ as
\eq{
J_{ab} = -\frac{1}{4} (\mb T_a \mb T_b - \mb T_b \mb T_a) = -\half 1 (\mb T_a \mb T_b - \eta_{ab})
}
so that $J_{aa} = 0$ and $J_{ab} = -J_{ba} = -\half 1 \mb T_a \mb T_b$ for $a \neq b$.

It can be proven that the scalar product $q = \Psi^* \eta$ is a Killing form for $\spin(r,s)$, that is
\eq{
q(X, [Y, Z]) = q([X,Y], Z), \quad \forall X, Y, Z \in \spin(r,s)
}

Now consider $\R^n$ with the standard euclidean metric $\delta$ and $\delta$-orthonormal basis $\{ T_1, \dots, T_n \} = \{ T_i \}$, and $\R^{n+1}$ with the standard lorentzian metric $\eta$ and $\eta$-orthonormal basis $\{ T_0, \dots, T_n \} = \{ T_0, T_i \} = \{ T_a \}$. Then the isometric immersion
\eq{
\mapp{\R^n}{\R^{n+1}}{(x^1, \dots, x^n)}{(0, x^1, \dots, x^n)}
}
induces, by functioriality, immersions of the Clifford algebras, orthogonal groups, spin groups, and spin algebras. In particular
\eq{
\Spin(n) &= \Spin(n,0) = \{ S \in \Spin_0(n,1) : S = \mb v_1 \cdots \mb v_{2k}, v_i \in \R^n \}
	\\
\spin(n) &= \spin(n,0) = \{ X \in \spin(n,1) : X = \mb x \mb y, x, y \in \R^n \}
}
and
\eq{
\ell(\Spin(n)) = \SO(n) \subset \SO(n,1)
}

One of the main results in \cite{orizz_bi_conn} is that whenever $n \geq 3$ the pair $(\Spin(n,1), \Spin(n))$ is a reductive pair. First recall that

\begin{defn}[Reductive Pair]
A pair of Lie groups $(G, H)$, with Lie algebras $\lie g$ and $\lie h$ respectively, is a reductive pair if:
\begin{enumerate}[(i)]
\item $H \subset G$ is a closed Lie subgroup, so that $\lie h$ is a subalgebra of $\lie g$;

\item there exists a reductive splitting, that is
\eq{
\lie g = \lie h \oplus \lie m \quad{\text{and}} \quad \Ad_G(H) \lie m \subset \lie m
}
where $\Ad_G(H)$ is the adjoint representation of $G$ restricted to $H$.
\end{enumerate}
\end{defn}

The theorem above then is
\begin{teo}
\label{spin_su_red_pair}
If $n > 3$, the pair $(\Spin_0(n, 1); \Spin(n))$ has a unique reductive splitting
\eq{
\spin(n, 1) = \spin(n) \oplus \lie m_0
}
If $n = 3$, we have a $1$-parameter family of reductive splittings
\eq{
\spin(3, 1) = \spin(3) \oplus \lie m_\beta
}
The parameter $\beta$ is called the \emph{Immirzi parameter} of the splitting. If we choose the basis
\eq{
L_k = \half 1 \tau_k = \half 1 \epsilon^{ij}{}_k \, J_{ij}
}
for $\su(2) = \spin(3)$, then
\eq{
H_k^{(\beta)} 
&= K_k - \beta L_k
	\\
&= J_{0k} - \beta L_k
}
is a basis for $\lie m_\beta$.
\end{teo}

\begin{rem}
One can verify that the generators for the case $n = 3$ satisfy the following commutation relations
\eq{
&[L_i, L_j] = \epsilon_{ij}{}^k \, L_k
	\\
&[L_i, H^{(\beta)}_j] = \epsilon_{ij}{}^k \, H^{(\beta)}_k
	\\
&[H^{(\beta)}_i, H^{(\beta)}_j] = -\epsilon_{ij}{}^k \[ (1 + \beta^2) L_k + 2\beta H^{(\beta)}_k \]
}

\end{rem}

\subsection{Spin Frames, Solder Form, and Bundle Reduction}

We now define spin frames on an $m$-dimensional manifold $M$ which admits metrics of signature $(r,s)$. There are topological obstructions both to the existence of metrics of a given signature and to the existence of spin frames, from here onwards we will be assuming that the manifold $M$ under consideration admits such structures. Refer to \cite{orizz_bi_conn} and references therein for a more detailed discussion.

\begin{defn}[Spin Frame]
\label{spin_struc_spin_frame}

Denote by $\ell \colon \Spin_0(r,s) \rarr \SO_0(r,s)$ the twofold covering of the orthogonal group by the relative spin group and by $i \colon \SO_0(r,s) \rarr \GL(m)$ the canonical embedding of the orthogonal group into the general linear group. A \emph{spin frame on $M$} is a pair $(Q, e)$ where $\sv Q = (Q, p, M, \Spin_0(r,s))$ is a $\Spin_0(r,s)$-bundle on $M$ and $e \colon Q \rarr L(M)$ is a principal bundle map, that is a commutative diagram
\commd{
Q \dar{p} \rar{e} \& L(M) \dar{\pi}
	\\
M \rar[equal] \& M
	\\
\Spin_0(r,s) \rar{i \circ \ell} \& \GL(m)
}
The commutative diagram above denotes that for any $S \in \Spin(r,s)$ and any $q \in Q$ we have
\eq{
e(q \. S) = e(q) \. (i \circ \ell)(S)
}
\end{defn}

Any spin frame $e$ on $M$ defines a metric and, if $M$ is orientable, it also induces an orientation and a volume form on $M$. The induced metric $g_e$ is defined as
\eq{
(g_e)_x (e(q)_a, e(q)_b) = \eta_{ab}, \quad \forall q \in Q_x
}
That is, the induced metric $g_{Q}$ is the metric for which the frames in the image of the spin frame $e$ are $\eta$-orthonormal (recall that the frames are \emph{ordered} bases). For any $x \in M$ and any $q, q' \in Q_x$ we have that there exists a unique $S \in \Spin_0(r,s)$ such that
\eq{
q' = q \. S \iff e(q') = e(q) \. (i \circ \ell)(S)
}
Since $\det ( (i \circ \ell)(S) ) = 1$, the frames in the image of the spin frame $e$ have the same orientation, and given that $e$ is a smooth map this defines an orientation on all of $M$. We then define the positively oriented frames as the frames in the image of the spin frame $e$. As for the volume form, consider the volume form $\nu_e$ induced by the metric $g_e$, this is defined by its action on the positively oriented, $g_e$-orthonormal frames
\eq{
(\nu_{e})_x (e(q)_1, \dots, e(q)_m) = 1, \quad \forall q \in Q_x
}

One can describe a spin frame also via a differential $1$-form on $Q$ which is valued in $\R^m$ called the \emph{solder form (for the spin frame $e$)} and is denoted by $\theta_e$. To define it first consider now the representation of the spin group $\Spin_0(r,s)$ on $\R^m$ via
\eq{
\mapp{\Spin_0(r,s) \times \R^m}{\R^m}{(\hat S, x)}{\( (i \circ \hat \ell)(\hat S) \) x}
}
where we denoted the action of $\GL(m)$ on $\R^m$ simply by juxtaposition. With some abuse of notation, we denote this representation simply by $\ell \colon \Spin_0(r,s) \ract \R^m$ and define the associated vector bundle $\sv E = Q^\ell = (E, p^\lambda, M, \R^m)$. The elements in $E$ are equivalence classes of pairs $(q, a) \in Q \times \R^m$ with
\eq{
(q', a') \sim (q, a) \iff \exists S \in \Spin_0(r,s) : \sys{
q' = q \. S
	\\
a' = \ell(\inv S) a
}
}
We denote the equivalence classes by $[q, a]_{\Spin_0(r,s)}$. Now recall that a frame $e_x \in L_x(M)$ can be regarded as an invertible linear map
\eq{
e_x \colon \R^m \rarr T_x M
}
The frame in $T_x M$ is the image of the standard frame $\{T_a \}_{a = 1, \dots, m}$ in $\R^m$. Now, for any $q \in Q_x$ and any vector $\Xi_q \in T_q Q$, we have the vector $T_q p(\Xi_q) \in T_x M$, so that $\inv{e(q)}(T_q p(\Xi_q))$ is an element of $\R^m$. Using this we can define the solder form $\theta_e$ as
\eq{
\map{\theta_e}{TQ}{\R^m}{\Xi_q}{\inv{e(q)}(T_q p(\Xi_q))}
}

Recall (see, for instance, \cite{kn1} p.\ 75) that given a principal $G$-bundle $\sv P = (P, \pi, M, G)$ and a representation $\lambda \colon G \rarr \GL(V)$ on a vector space $V$, a $V$-valued differential $k$-form $\alpha$ on $P$ is \emph{pseudotensorial of type $(\lambda, V)$}, written as $\alpha \in \Omega^k(P, V)^\lambda$, if
\eq{
(R_g)^* \alpha = \lambda(\inv g) \alpha
}
where $R_g$ is the right action of $g \in G$ on $P$. If $\alpha$ is horizontal, that is it vanishes on vertical vectors, then $\alpha$ is \emph{tensorial of type $(\lambda, V)$} and we write $\alpha \in \Omega_H^k(P, V)^\lambda$.

We now show that the solder form $\theta_e$ is tensorial of type $(\ell, \R^m)$, that is $\theta_e \in \Omega^1_H(Q, \R^m)^\ell$. Since vertical vectors $\Xi_q \in V_q Q$ satisfy $T_q p(\Xi_q) = 0$ we have that $\theta_e$ is horizontal. Since the right action $R_S$ satisfies $p \circ R_S = p$ for any $S \in \Spin_0(r,s)$ we get
\eq{
((R_S)^* \theta_e)(\Xi_q)
&= \theta_e(T_q R_S (\Xi_q))
	\\
&= \inv{e(q \. S)}(T_{q \. S} p(T_q R_S(\Xi_q)))
	\\
&= \inv{e(q \. S)}(T_{q} (p \circ R_S)(\Xi_q))
	\\
&= \inv{e(q \. S)}(T_{q} p (\Xi_q))
	\\
&= \ell(\inv S) \( \inv{e(q)}(T_{q} p (\Xi_q)) \)
	\\
&= \ell(\inv S) (\theta_e(\Xi_q) )
}
By arbitrarity of $\Xi_q$ we get $(R_S)^* \theta_e = \ell(\inv S) \theta_e$, which is the definition equivariance. 

We now use the solder form $\theta_e$ to prove that the tangent bundle $TM$ and the vector bundle $E$ are isomorphic \emph{as vector bundles}. In fact for any vector $v_x \in T_x M$ and any $q \in Q$, choose a preimage $\Xi_q \in T_x Q$ and define the map $\theta_e$
\eq{
\map{\theta_e}{TM}{E}{v_x}{[q, \theta_e(\Xi_q)]_{\Spin_0(r,s)}}
}
The definition is well-given by the properties of the solder form. This bundle map is injective by definition and by rank considerations it is also surjective, hence it is a vector bundle isomorphism.

Consider now the lorentzian case $(r,s) = (m-1,1) = (n,1)$. Since the pair $(\Spin(n,1), \Spin(n))$ is a reductive pair one could define a $\spin(n)$-valued connection out of any $\spin(n,1)$-connection $\omega$ by taking its $\spin(n)$-part in the reduction. This procedure relies on the following theorem.

\begin{teo}\label{princ_bun_red_conn}
Let $(G, H)$ be a reductive pair with Lie algebras $\lie{g}, \lie{h}$ respectively and reductive splitting $\lie g = \lie h \oplus \lie m$. Consider a principal $G$-bundle $\sv P = (P, \pi, M, G)$ with an $H$-reduction $\sv Q = (Q, \pi', M, H)$, which is a principal bundle morphism
\commd{
Q \dar{\pi'} \rar[hook]{\iota} \& P \dar{\pi}
	\\
M \rar[equal] \& M
	\\
H \rar{i} \& G
}
Given a principal connection $\omega$ on $P$, call $A$ its $\lie{h}$-component and $\kappa$ its $\lie{m}$-component. Then the $\lie h$-part $A \in \Omega^1(P, \lie h)^{\Ad}$ when restricted to $Q$ is a principal connection on $Q$.
\end{teo}

To use this result in the case of a spin frame $(Q,e)$ we need to verify that there exists at least one reduction from $\Spin_0(n,1)$ to $\Spin(n)$. In principle there are topological obstructions to the existence of a structure group reduction, but in \cite{orizz_bi_conn} we proved that for a spin frame $(Q,e)$ in lorentzian signature there always exists a principal $\Spin(n)$-bundle ${}^+ \sv Q = ({}^+ Q, p', M, \Spin(n))$ which is a reduction. Therefore we can use \cref{spin_su_red_pair} to split any $\spin(n,1)$-connection $\omega$ into its $\spin(n)$-part $A$ and its $\lie m$- (or $\lie m_\beta$-) part $\kappa$. Restricting $A$ to ${}^+ Q$ gives a $\spin(n)$-connection on ${}^+ Q$.

\subsection{Spin Connection, Torsion and Contorsion}

The subject of principal connections is well-known and can be found, for example, in the book by Kobayashi and Nomizu \cite{kn1}. The special case of orthogonal/spin connections and their classification via torsion and contorsion tensors can be found in the book by Bleecker \cite{bleecker_gauge} (p.\ 77 and onwards).

We recall the definition of vector-valued forms on a principal $G$-bundle $P$.

\begin{defn}[$V$-Valued $k$-forms on $P$]
Consider a principal $G$-bundle $\sv P = (P, \pi, M, G)$ and a vector space $V$. The vector space of maps
\eq{
\alpha \colon \Lambda^k TP \rarr V
}
is denoted by $\Omega^k(P, V)$ and is the set of $V$-valued $k$-forms on $P$. The $V$-valued $k$-forms which vanish on vertical vectors $VP$ are called horizontal and form a subspace denoted by $\Omega^k_H(P, V)$.
%Since $V$ is a finite dimensional vector space, any $\alpha \in \Omega^k(P, V)$ can be written as a finite linear combination of decomposable elements
\eq{
\alpha = \pphi \ot v, \quad \pphi \in \Omega^k(P), v \in V
}

If $\lambda \colon G \rarr \GL(V)$ is a representation of $G$ on $V$ we define the \emph{$V$-valued $k$-forms on $P$ of type $\lambda$} (or \emph{$k$-forms on $P$ of type $(\lambda, V)$}) as the $\alpha \in \Omega^k(P, V)$ such that
\eq{
(R_g)^* \alpha(p \. g) = \lambda(\bar g) \. \alpha(p), \quad \forall p \in P, \forall g \in G
}
and denote the vector space of such forms as $\Omega^k(P, V)^\lambda$. A $V$-valued $k$-form of type $\lambda$ which vanishes on vertical vectors $VP$ is called \emph{tensorial}, the vector space of such forms is denoted by $\Omega^k_H(P, V)^\lambda$.
\end{defn}

It is a known fact (see, for instance, \cite{kn1} p.\ 77) that principal connections on a principal $G$-bundle $\sv P = (P, \pi, M, G)$ are pseudotensorial $1$-forms on $P$ of type $(\Ad, \lie g)$. With the choice of a principal connection we can define the exterior covariant differential.

\begin{defn}[Exterior Covariant Differential]
\label{princ_bun_ext_cov_d}
Consider a principal $G$-bundle $\sv P = (P, \pi, M, G)$, a representation $\lambda \colon G \rarr \GL(V)$ of $G$ on a vector space  $V$, and the relative associated vector bundle bundle $\sv P^\lambda = \sv E = (E, \pi^\lambda, M, V)$. For a fixed principal connection $\omega$ on $P$ denote by $h \colon TP \rarr HP$ the horizontal projector. The \emph{exterior covariant differential} of a $V$-valued $k$-form $\alpha$ on $P$ is defined as
\eq{
(\D\omega \alpha)(\Xi_0, \dots, \Xi_k) = d \alpha (h\Xi_0, \dots, h \Xi_k), \quad \Xi_i \in TP
}
As such it is a linear map
\eq{
\map{\D\omega}{\Omega^k(P, V)}{\Omega^{k+1}_H(P, V)}{\alpha}{\D\omega \alpha}
}
For any representation $\lambda \colon G \rarr \GL(V)$ of $G$ on a vector space $V$, the exterior covariant differential satisfies 
\eq{
& \D\omega(\Omega^k(P, V)^\lambda) \subset \Omega^{k+1}_H(P, V)^\lambda
}
That is, it maps pseudotensorial forms to tensorial forms.

For tensorial forms $\alpha \in \Omega^k_H(P, V)^\lambda$ we have the explicit formula
\eq{
\D\omega \alpha = d \alpha + T\lambda(\omega) \w \alpha
}
where $T\lambda \colon \lie g \rarr \GL(V)$ is the Lie algebra representation induced by $\lambda$ and 
\eq{
(T\lambda(\omega) \w \alpha)(\Xi_0, \dots, \Xi_k) 
&= \frac{1}{k!} \sum_{\sigma \in \lie S_{k+1}} \sgn \sigma \, T\lambda \( \omega(\Xi_{\sigma(0)}), \alpha(\Xi_{\sigma(1)}, \dots, \Xi_{\sigma(k)}) \)
}
\end{defn}

For the special case of a spin frame $(Q, e)$ on $M$, with solder form $\theta_e \in \Omega^1_H(Q, \R^m)^\ell$ and a given spin connection $\omega \in \Omega^1(Q, \spin(r,s))^{\Ad}$, we define the following objects:
\begin{itemize}
\item the \emph{curvature/field-strength $R_\omega$ of $\omega$}
\eq{
R_\omega = \D\omega \omega
}
The curvature is a tensorial $2$-form of type $(\Ad, \spin(r,s))$, $R_\omega \in \Omega^2_H(Q, \spin(r,s))^{\Ad}$.

\item the \emph{torsion $\Theta_\omega$ of $\theta_e$ with respect to $\omega$}
\eq{
\Theta_\omega = \D\omega \theta_e
}
The torsion is a tensorial $2$-form of type $(\ell, \R^m)$, $\Theta_\omega \in \Omega^2_H(Q, \R^m)^\ell$ 

\item for any other spin connection $\omega'$ we define the \emph{contorsion $C_{\omega', \omega}$ of $\omega'$ relative to $\omega$}
\eq{
C_{\omega', \omega} = \omega' - \omega
}
The contorsion is a tensorial $1$-form of type $(\Ad, \spin(r,s))$, $C_{\omega', \omega} \in \Omega^1_H(Q, \spin(r,s))^{\Ad}$
\end{itemize}

Notice that we can use the contorsion $C_{\omega', \omega}$ to express the torsion $\Theta_{\omega'}$ with respect to $\Theta_\omega$, in fact
\eq{
\Theta_{\omega'} 
&= \D{\omega'} \theta_e
	\\
&= d\theta_e + T\ell(\omega') \w \theta_e
	\\
&= d\theta_e + T\ell(\omega + C_{\omega', \omega}) \w \theta_e 
	\\
&= \Theta_\omega + T\ell(C_{\omega', \omega}) \w \theta_e
}

The simplest possible case would be that of a connection $\omega$ with zero torsion $\Theta_\omega = 0$, which one could use as the origin in the affine space of spin connections. As in the context of metric manifolds, a torsionless connection always exists and it is unique: it is the \emph{Levi--Civita Connection $\{ e \}$ of the spin frame $e$}.

\begin{teo}[Levi--Civita Connection of a Spin Frame $e$]
\label{spin_struc_levi_civita}
For a fixed spin frame $(Q, e)$ there exists a unique spin connection $\{ e \}$ that is torsionless, it is called \emph{Levi--Civita connection of $e$}. All other spin connections $\omega$ are obtained via
\eq{
\omega = \{ e \} + C_\omega
}
where $C_\omega$ is a tensorial $1$-form of type $(\Ad, \spin(r,s))$ on $Q$ know as the \emph{contorsion tensor of $\omega$} and is unique. The torsion $\Theta_\omega$ of $\omega$ is given by
\eq{
\Theta_\omega = \D\omega \theta_e = T\ell(C_\omega) \w \theta_e
}
\end{teo}

\begin{cor}
The correspondence between spin connections and torsion/contorsion forms is one-to-one, that is
\eq{
\omega = \omega' \iff \Theta_\omega = \Theta_{\omega'} \iff C_\omega = C_{\omega'}
}
\begin{proof}
The implications starting from $\omega = \omega'$ and $C_\omega = C_{\omega'}$ are a direct consequence of the definitions. Suppose, then, that we have two spin connections $\omega, \omega'$ with the same torsion $\Theta_\omega = \Theta_{\omega'}$. Since
\eq{
0 = \Theta_{\omega'} - \Theta_{\omega} = T\ell(C_{\omega'} - C_\omega) \w \theta_e
}
Then the connection $\{ e \} + C_{\omega'} - C_{\omega}$ is torsionless, but the unique torsionless connection is $\{ e \}$ itself, meaning that
\eq{
C_{\omega'} - C_{\omega} = 0 \iff C_{\omega'} = C_{\omega} \iff \omega' = \omega
}
which is the thesis.
\end{proof}
\end{cor}

%
%% %%
\section{Calculus for $\Lambda \R^m$-valued forms on $Q$}\label[sec]{vec_forms_vec_forms}

From here onwards we consider a spin frame $e \colon Q \rarr L(M)$, where $Q$ is a spin bundle $\sv Q = (Q, p, M, \Spin_0(r,s))$ on $M$, as such we denote the solder form $\theta_e$ simply by $\theta$. As before we have the group morphisms
\commd{
\Spin_0(r,s) \rar[two heads]{\ell} \& \SO_0(r,s) \rar[hook]{i} \& \GL(m)
}
where $\ell \colon \Spin_0(r,s) \surarr \SO_0(r,s)$ is the twofold covering map of the spin group (this is the universal covering for $(r,s) = (m,0)$ or $(r,s) = (m-1,1)$), and $i \colon \SO_0(r,s) \injrarr \GL(m)$ is the standard inclusion of matrix groups. Then the spin group $\Spin_0(r,s)$ acts on $\R^m$ via
\eq{
\mapp{\Spin_0(r,s) \times \R^m}{\R^m}{(\hat S, x)}{\( (i \circ \hat \ell)(\hat S) \) x}
}
where we denoted the action of $\GL(m)$ on $\R^m$ simply by juxtaposition. With some abuse of notation, we denote this representation simply by $\ell \colon \Spin_0(r,s) \ract \R^m$ and define the associated vector bundle $\sv E = Q^\ell = (E, p^\lambda, M, \R^m)$ which is isomorphic \emph{as vector bundles} to $TM$ via the solder form $\theta$ of $Q$.

\begin{defn}[$\Lambda \R^m$-Valued Forms on a $\Spin_0(r,s)$-Bundle]

Consider a spin bundle $\sv Q = (Q, p, M, \Spin_0(r,s))$ on $M$ and the standard representation $\ell \colon \Spin_0(r,s) \rarr \SO_0(r,s)$ on $\R^m$. We define the vector space of \emph{$\Lambda^h \R^m$-valued $k$-forms on $Q$} as
\eq{
\Omega^{k,h}(Q, \R^m) = \Omega^k(Q) \ot \Lambda^h \R^m = \{ \Phi \colon \Lambda^k TQ \rarr \Lambda^h \R^m \}
}
Differential forms $\Phi \in \Omega^{k,h}(Q, \R^m)$ are \emph{homogeneous forms of bidegree $\deg \Phi = (k,h)$}. We define the direct sums
\eq{
&\Omega^{\bullet, h}(Q, \R^m) = \bigoplus_{k = 1}^{\dim Q} \Omega^{k, h}(Q, \R^m)
	\\
&\Omega^{k, \bullet}(Q, \R^m) = \bigoplus_{h = 1}^{m} \Omega^{k, h}(Q, \R^m)
	\\
&\Omega(Q, \R^m) = \bigoplus_{k = 1}^{\dim Q} \Omega^{k, \bullet}(Q, \R^m) = \bigoplus_{h = 1}^{m} \Omega^{\bullet, h}(Q, \R^m)
}

Due to the fact that $\Lambda^k \R^m$ is a finite-dimensional vector space, a generic $\Phi \in \Omega^{k,h}(Q, \R^m)$ is a finite linear combination of \emph{decomposable elements}, that is forms of the type
\eq{
\Phi = \pphi \ot v ,\quad \text{with } \pphi \in \Omega^k(Q), v \in \Lambda^h \R^m
}

If we consider the actions $\Lambda^h \ell \colon \Spin_0(r,s) \ract \Lambda^h \R^m$ induced by functoriality, we can define \emph{$k$-forms on $Q$ of type $(\Lambda^h \ell, \Lambda^h \R^m)$} and we denote them by 
\eq{
\Omega^{k,h}(Q, \ell) = \Omega^{k,h}(Q, \Lambda^h\R^m)^{\Lambda^h \ell}
}
The equivariance condition is
\eq{
(R_S)^* \Phi = (\Lambda^h \ell(\inv S)) (\Phi), \quad \forall \Phi \in \Omega^{k,h}(Q, \ell), \forall S \in \Spin_0(r,s)
}
On a decomposable element $\Phi = \pphi \ot v$ we have
\eq{
\( \Lambda^h \ell(S) \)(\Phi) = \pphi \ot (\Lambda^h \ell(S))(v)
}
As before, forms of type $(\Lambda^h \ell, \Lambda^h \R^m)$ are also called \emph{pseudotensorial}. Forms $\Phi \in \Omega^{k,h}(Q, \ell)$ which vanish on vertical vectors are denoted by $\Omega^{k,h}_H(Q, \ell)$ and are called \emph{tensorial}.

The spaces $\Omega^{\bullet, h}(Q, \ell), \Omega^{k, \bullet}(Q, \ell)$ and $\Omega(Q, \ell)$ and their tensorial subspaces are defined as above.
\end{defn}

The main reason for the unifying formalism is that, by using the isomorphism $\spin(r,s) \simeq \Lambda^2 \R^m$, we can describe all of the spin frame-related objects using the same language:
\eq{
&\text{solder form } \theta \in \Omega^{1,1}_H(Q, \ell)
	\\
&\text{spin connection } \omega \in \Omega^{1,2}(Q, \ell)	
	\\
&\text{contorsion form } C_\omega \in \Omega^{1,2}_H(Q, \ell)	
	\\
&\text{torsion form } \Theta_\omega \in \Omega^{2,1}_H(Q, \ell)	
	\\
&\text{curvature form } R_\omega \in \Omega^{2,2}_H(Q, \ell)		
}

We can extend various operations defined on $\Omega(Q)$ and $\Lambda \R^m$ to $\Omega^{k,h}(Q, \R^m)$, starting with the wedge product:

\begin{defn}[Kulkarni--Nomizu Product]
The space $\Omega(Q, \R^m)$ is a bigraded algebra. The \emph{Kulkarni--Nomizu (KN) product $\ow$} is defined on decomposable elements as
\eq{
\map{\ow}{\Omega^{k,h}(Q, \R^m) \times \Omega^{k',h'}(Q, \R^m)}{\Omega^{k+k',h+h'}(Q, \R^m)}{(\Phi = \pphi \ot v, \Psi = \psi \ot w)}{(\pphi \w \psi) \ot (v \w w)}
}
and extended by linearity to all elements. The KN product is a bigraded derivation, for $\Phi \in \Omega^{k,h}(Q, \R^m)$ and $\Psi \in \Omega^{k',h'}(Q, \R^m)$ we have
\eq{
\Phi \ow \Psi = (-1)^{kk'}(-1)^{hh'} \Psi \ow \Phi
}
which descends from the analogous relations for wedge products in $\Omega(Q)$ and $\Lambda \R^m$. Notice that the subspaces of pseudotensorial forms $\Omega(Q, \ell)$ and of tensorial forms $\Omega_H(Q, \ell)$ are subalgebras with the induced KN product $\ow$.
\end{defn}

\begin{defn}[Exterior Differential on $\Omega(Q, \R^m)$]
The exterior differential $d \colon \Omega^k(Q) \rarr \Omega^{k+1}(Q)$ extends to $\Omega^{k,h}(Q, \R^m)$. On decomposable elements we have
\eq{
\map{d}{\Omega^{k,h}(Q, \R^m)}{\Omega^{k+1, h}(Q, \R^m)}{\Phi = \pphi \ot v}{d\Phi = (d\pphi) \ot v}
}
The Leibniz formula in this case has the form, for $\Phi \in \Omega^{k,h}(Q, \R^m)$ and $\Psi \in \Omega^{k',h'}(Q, \R^m)$
\eq{
d(\Phi \ow \Psi)
%&= d(\pphi \w \psi ) \ot (v \w w) 
%	\\
%&= \[ d \pphi \w \psi + (-1)^k \pphi \w d \psi \] \ot (v \w w)
%	\\
&= d \Phi \ow \Psi + (-1)^k \, \Phi \ow d \Psi
}
The exterior differential $d$ restricts to a derivation on the subalgebra of pseudotensorial forms $\Omega(Q, \ell)$ \uline{but not} on the subalgebra of tensorial forms $\Omega_H(Q, \ell)$.
\end{defn}

The definition of torsion $\Theta_\omega$ and curvature $R_\omega$ suggest the following. For any representation $\rho \colon \lie g \rarr \gl(m)$ of a Lie algebra $\lie g$ we have, by functoriality, the induced actions $\Lambda^h \rho$ on $\Lambda^h \R^m$, therefore we can extend $\rho$ to $\Omega(Q, \R^m)$. As before the action is completely determined by its effect on decomposable elements, for $X \in \lie g$, $\pphi \in \Omega^k(Q)$ and $v \in \Lambda^h \R^m$ we then have
\eq{
\rho(X) (\pphi \ot v) = \pphi \ot (\Lambda^h \rho(X) (v))
}
Using this fact we can define the action of any $\lie g$-valued $q$-form in $\Omega^q(Q, \lie g)$ on $\Omega^{k,h}(Q, \R^m)$, it suffices to describe this action on decomposable elements: consider $\alpha \in \Omega^q(Q)$, $X \in \lie g$, $\pphi \in \Omega^k(Q)$, and $v \in \Lambda^h\R^m$
\eq{
\map{\Lambda^h \rho}{\Omega^q(Q, \lie g) \times \Omega^{k,h}(Q, \R^m)}{\Omega^{k+q, h}(Q, \R^m)}{(\alpha \ot X , \pphi \ot v)}{(\alpha \w \pphi) \ot \Lambda^h \rho(X)(v)}
}

If the action $\rho$ is valued in $\so(r,s) \subset \gl(m)$ then we have that all actions $\Lambda^h \rho$ can be restricted to pseudotensorial forms $\Omega(Q, \ell)$ but not to tensorial forms $\Omega_H(Q, \ell)$.

\begin{rem}
We also write this action as $\Lambda^h\rho(\alpha \ot X) \w (\pphi \ot v)$. In general if $\alpha \in \Omega^q(Q, \lie g)$ and $\Phi \in \Omega^{k,h}(Q, \R^m)$ we have that
\eq{
&(\Lambda^h\rho(\alpha) \w \Phi)(\Xi_1, \dots, \Xi_q, \Xi_{q+1}, \dots, \Xi_{q + k})  = 
	\\
&\qquad \qquad = \frac{(q + k)!}{q! \, k!} \sum_{\sigma \in \lie S_{q+k}} \sgn \sigma \, \Lambda^h\rho\( \alpha (\Xi_{\sigma(1)}, \dots, \Xi_{\sigma(q)}) \)(\Phi(\Xi_{\sigma(q+1)}, \dots, \Xi_{\sigma(q+k)}))
}
\end{rem}

The case of most interest for us is when $\lie g = \spin(r,s)$ and $\rho$ is the standard representation of $\spin(r,s)$ on $\R^m$ induced by the action of $\Spin_0(r,s)$, that is $\rho = T\ell$. In this situation we simplify by introducing the dot notation: for any $X \in \spin(r,s)$ and $\Phi \in \Omega^{k,h}(Q, \R^m)$ we have
\eq{
X \cw \Phi = \Lambda^h T\ell(X)(\Phi)
}

Using dot notation we can treat exterior covariant derivatives on forms of different bidegree in a unified way. We adapt \cref{princ_bun_ext_cov_d} to this case.

\begin{defn}[Exterior Covariant Differential on $\Omega^{k,h}(Q, \ell)$]
\label{vec_forms_ext_cov_d_spin}
Consider a spin bundle $\sv Q = (Q, p, M, \Spin_0(r,s))$ on $M$ and the standard representation $\ell \colon \Spin_0(r,s) \rarr \SO_0(r,s)$ on $\R^m$. For a fixed spin connection $\omega$ on $Q$ denote by $h \colon TQ \rarr HQ$ the horizontal projector. The \emph{exterior covariant differential} of a $\Lambda \R^m$-valued $k$-form $\Phi$ on $Q$ is defined as
\eq{
(\D\omega \Phi)(\Xi_0, \dots, \Xi_k) = d \Phi (h\Xi_0, \dots, h \Xi_k), \quad \Xi_i \in TQ
}
As such it is a linear map
\eq{
\map{\D\omega}{\Omega^{k,h}(Q, \R^m)}{\Omega^{k+1}_H(Q, \R^m)}{\Phi}{\D\omega \Phi}
}
Exterior covariant differentiation preserves pseudotensoriality and tensoriality, that is $\D\omega \(\Omega^{k,h}(Q, \ell) \) \subset \Omega^{k+1,h}_H(Q, \ell)$. 

We know restate some of the known properties satisfied by $\D\omega$ using our notation
\begin{enumerate}[(i)]
\item for $\Phi \in \Omega^{k,h}_H(Q, \R^m)$ we have
\eq{
\D\omega \Phi = d \Phi + \omega \cw \Phi
}

\item the \emph{curvature form $R_\omega$} of $\omega$ is
\eq{
R_\omega = \D\omega \omega 
&= d\omega + \half 1 \omega \cw \omega
	\\
&= d\omega + \half 1 [\omega \w \omega]
}
where $[\omega \w \omega] = \ad(\omega) \w \omega$, with $\ad \colon \spin(r,s) \ract \spin(r,s)$ the adjoint representation of the Lie algebra on itself;

\item the \emph{Bianchi identity} is
\eq{
\D\omega R_\omega = \D\omega{}^2 \omega = 0
}
For the solder form $\theta \in \Omega^{1,1}_H(Q, \R^m)$ of $Q$ the \emph{Bianchi identity} has the form
\eq{
\D\omega \Theta_\omega = \D\omega{}^2 \theta = R_\omega \cw \theta
}
\end{enumerate}
\end{defn}

\begin{rem}
Notice that for $\Lambda^2 T\ell = \ad \colon \spin(r,s) \ract \spin(r,s)$ will still use the commutator notation $[\. \w \. ]$, mostly because it is a very common in literature.
\end{rem}

We end this section by characterizing the interaction between $\D\omega, \ow$, and $\cw$.
\begin{prop}
We have:

\begin{enumerate}[(i)]
\item for $\Phi \in \Omega^{k,h}(Q, \R^m)$ and $\Psi \in \Omega^{k',h'}(Q, \R^m)$ then
\eq{
\D\omega (\Phi \ow \Psi) 
%&= \D\omega((\pphi \ot v) \ow (\psi \ot w))
%	\\
%&= \D\omega((\pphi \w \psi) \ot (v \w w))
%	\\
%&= ( \D\omega\pphi \w \psi + (-1)^{kk'} \phi \w \D\omega \psi ) \ot (v \w w)
%	\\
&= \D\omega \Phi \ow \Psi + (-1)^{kk'} \Phi \ow \D\omega \Psi
}

\item given a Lie algebra representation $\rho \colon \lie g \rarr \gl(m)$, $\Theta \in \Omega^q(Q, \lie g)$, and $\Phi \in \Omega^{k,h}(Q, \R^m)$ then
\eq{
\D\omega(\rho(\Theta) \w \Phi) 
%&= \D\omega (\rho(\alpha \ot X) \w (\pphi \ot v))
%	\\
%&= \D\omega \( (\alpha \w \pphi) \ot ( \rho(X) v) \)
%	\\
%&= \D\omega ((\alpha \w \pphi)) \ot ( \rho(X) v)
%	\\
%&= \( \D\omega \alpha \w \pphi + (-1)^q \alpha \w \D\omega \pphi\) \ot ( \rho(X) v)
%	\\
&= \rho\( \D\omega (\Theta) \) \w \Phi + (-1)^q \rho(\Theta) \w \D\omega \Phi
}

\item given a Lie algebra representation $\rho \colon \lie g \rarr \gl(m)$, $\Theta \in \Omega^q(Q, \lie g)$, and forms $\Phi \in \Omega^{k,h}(Q, \R^m)$ and $\Psi \in \Omega^{k',h'}(Q, \R^m)$ then
\eq{
(\rho(\Theta) \w \Phi) \ow \Psi
%&= (\rho(\alpha \ot X) \w (\pphi \ot v)) \ow (\psi \ot w)
%	\\
%&= \( (\alpha \w \pphi) \ot (\rho(X) v) \) \ow (\psi \ot w)	
%	\\
%&= \( (\alpha \w \pphi \w \psi) \ot ((\rho(X) v) \w w) \) 
%	\\
%&= \( (\alpha \w \pphi \w \psi) \ot ((\rho(X) (v \w w) - v \w (\rho(X) w)) \) 	
%	\\
%&= \rho(\Theta) \w (\Phi \ow \Psi) - (-1)^{qk} (\pphi \w \alpha \w \psi) \ot (- v \w (\rho(X) w))	
%	\\
&= \rho(\Theta) \w (\Phi \ow \Psi) + (-1)^{qk} \Phi \ow ( \rho(\Theta) \w \Psi)
}
\end{enumerate}
\begin{proof}
By linearity, it suffices to prove the formulas for decomposable $\Theta = \alpha \ot X$, $\Phi = \pphi \ot v$, and $\Psi = \psi \ot w$. Then
\begin{enumerate}[(i)]
\item 
\eq{
\D\omega (\Phi \ow \Psi) 
&= \D\omega((\pphi \ot v) \ow (\psi \ot w))
	\\
&= \D\omega((\pphi \w \psi) \ot (v \w w))
	\\
&= ( \D\omega\pphi \w \psi + (-1)^{kk'} \phi \w \D\omega \psi ) \ot (v \w w)
	\\
&= \D\omega \Phi \ow \Psi + (-1)^{kk'} \Phi \ow \D\omega \Psi
}

\item 
\eq{
\D\omega(\rho(\Theta) \w \Phi) 
&= \D\omega (\rho(\alpha \ot X) \w (\pphi \ot v))
	\\
&= \D\omega \( (\alpha \w \pphi) \ot ( \rho(X) v) \)
	\\
&= \D\omega ((\alpha \w \pphi)) \ot ( \rho(X) v)
	\\
&= \( \D\omega \alpha \w \pphi + (-1)^q \alpha \w \D\omega \pphi\) \ot ( \rho(X) v)
	\\
&= \rho\( \D\omega (\Theta) \) \w \Phi + (-1)^q \rho(\Theta) \w \D\omega \Phi
}

\item 
\eq{
(\rho(\Theta) \w \Phi) \ow \Psi
&= (\rho(\alpha \ot X) \w (\pphi \ot v)) \ow (\psi \ot w)
	\\
&= \( (\alpha \w \pphi) \ot (\rho(X) v) \) \ow (\psi \ot w)	
	\\
&= \( (\alpha \w \pphi \w \psi) \ot ((\rho(X) v) \w w) \) 
	\\
&= \( (\alpha \w \pphi \w \psi) \ot ((\rho(X) (v \w w) - v \w (\rho(X) w)) \) 	
	\\
&= \rho(\Theta) \w (\Phi \ow \Psi) - (-1)^{qk} (\pphi \w \alpha \w \psi) \ot (- v \w (\rho(X) w))	
	\\
&= \rho(\Theta) \w (\Phi \ow \Psi) + (-1)^{qk} \Phi \ow ( \rho(\Theta) \w \Psi)
}
\end{enumerate}

\end{proof}
\end{prop}

%%% %%%
\section{Tensorial Forms $\Omega_H(Q, \ell)$ and Hodge Operators}\label[sec]{vec_forms_hodge}

Recall that tensorial forms $\Omega^{k,h}_H(Q, \ell)$ are module-isomorphic to the subspace of $\Gamma(\Lambda^h E) \ot \Omega^k(M)$ (see, for instance, \cite{kn1} p.\ 76 ). We will always assume the isomorphism so that for a decomposable element $\pphi \ot v \in \Omega^k_H(Q, \ell)$, $\pphi$ can denote both the form in $\Omega^k_H(Q)$ and its corresponding form in $\Omega^k(M)$, analogously $v$ denotes both the element in $\Lambda^h \R^m$ and the section in $\Gamma(\Lambda^h E)$.

By definition, the metric $\eta$ on $\R^m$ is invariant under the action $\ell$ of $\Spin_0(r,s)$ on $\R^m$, using this we can define a metric $h$ of signature $(r,s)$ on $E$, the vector bundle associated to $Q$ via $\ell$. Explicitly it is
\eq{
h_x([q, a]_{\Spin_0(r,s)}, [q, b]_{\Spin_0(r,s)}) = \eta(a, b), \quad \forall q \in Q_x
}
Using that the solder form $\theta$ can be seen as an isomorphism between $TM$ and $E$, we get that $g = g_e = \theta^* h$, see \cite{kn1} for the details.

The metric $g$ defines two musical isomorphisms, the \emph{flat isomorphism $\flat_g$} and the \emph{sharp isomorphism $\sharp_g$}, which are inverses. The flat isomorphis $\flat_g$ is
\eq{
\map{\flat_g}{TM}{T^*M}{v_x}{(v_x)^{\flat_g} = g(v_x, -)}
}
This is an isomorphism since $g$ is non degenerate. Its inverse is the \emph{sharp isomorphism $\sharp_g$}
\eq{
\map{\sharp_g}{T^*M}{TM}{\alpha_x}{(\alpha_x)^{\sharp_g} = v_x \iff g(v_x, w_x) = \alpha_x(w_x), \quad \forall w_x \in T_xM}
}
Then the induced metric on $T^*M$ is again denoted by $g$ and is defined by
\eq{
\map{g_x}{T^*_x M \ot T^*_x M}{\R}{(\alpha_x, \beta_x)}{g_x(\alpha_x, \beta_x) = g\( (\alpha_x)^{\sharp_g}, (\beta_x)^{\sharp_g} \)}
}
One can also induce a metric on the spaces $\Lambda^k T^*M$, this is done by prescribing the metric on homogeneous elements. For $(\alpha_1)_x, \dots, (\alpha_k)_x, (\beta_1)_x, \dots, (\beta_k)_x \in T^*_x M$ we again denote the scalar product by $g$ and define it as
\eq{
&g_x \( (\alpha_1)_x \w \dots \w (\alpha_k)_x, (\beta_1)_x \w \dots \w (\beta_k)_x \) = 
	\\
&\quad = 
\begin{vmatrix} 
g\( (\alpha_1)_x, (\beta_1)_x \)  & \dots & g\( (\alpha_1)_x, (\beta_k)_x \)
	\\
\vdots & \ddots & \vdots
	\\	
g\( (\alpha_k)_x, (\beta_1)_x \)  & \dots & g\( (\alpha_k)_x, (\beta_k)_x \)
\end{vmatrix} = \det g\( (\alpha_i)_x, (\beta_j)_x \) 
}
In a similar manner one can extend the metric $h$ on $E$ to a metric on $\Lambda^h E$, which we will still denote by $h$.

Using that $\Lambda^2 E$ is the bundle associated to $Q$ via the induced action $\Lambda^2 \ell$ of $\Spin_0(r,s)$ on $\Lambda^2 \R^m$, we get that the metric $h$ on $\Lambda^2 E$ corresponds to the metric $\eta$ on $\Lambda^2 \R^m$ by the same construction which relates $h$ on $E$ to $\eta$ on $\R^m$. Using the isomorphism $\Psi \colon \spin(r,s) \rarr \Lambda^2 \R^m$ we defined the Killing form $q = \Psi^* \eta$ on $\spin(r,s)$. If we denote by $[ - , - ]$ the commutator induced on $\Lambda^2 E$ by the commutator/adjoint representation on $\Lambda^2 \R^m \simeq \spin(r,s)$, then we have again that for any $x \in M$
\eq{
h_x([u_x, v_x], w_x) = h_x(u_x, [v_x, w_x]), \quad \forall u_x,v_x,w_x \in \Lambda^2_x E
}

%\begin{prop}[Metric $g$ and Solder Form $\theta$]
%The metric $g$ on $M$ and the solder form $\theta$ are related by
%\eq{
%g_x = h_x(\theta_q, \theta_q), \quad \forall q \in Q_x
%}
%\begin{proof}
%From \cref{princ_bun_inv_tens} we have that
%\eq{
%g_x(v_x, w_x) = h_x(\theta(v_x), \theta(w_x)), \quad \forall v_x, w_x \in T_x M
%}
%where $\theta \colon TM \rarr E$ is the solder form. Under the correspondence $\Omega^1_H(Q, \ell) \lrarr \Omega^1(M, E)$ we then have that
%\eq{
%h_x(\theta(v_x), \theta(w_x)) = \eta(\theta_q(\Xi_q), \theta_q(\Xi'_q))
%}
%where $q \in \Q_x$ and $T_q p(\Xi_q) = v_x, T_q p(\Xi'_q) = w_x$
%\end{proof}
%\end{prop}

The manifold $M$ is orientable since we have a $\Spin_0(r,s)$-bundle on it, then also $E$ is orientable since it is isomorphic to $TM$. We can then define the two metric volume forms $\nu_g$ and $\nu_h$ on $TM$ and $E$ which are again related by the solder form
\eq{
\nu_g = \theta^* \nu_h
}
Denote by $n_h$ the section of $\Lambda^m E$ dual to $\nu_h$, meaning that it satisfies
\eq{
\nu_h(n_h) = 1
%h(\nu_h,  \nu_h)
}
Notice that $n_h$ is the section $n_h \colon M \rarr \Gamma(\Lambda^n E)$ which corresponds to the volume $m$-vector $n_\eta$ induced by $\eta$ on $\R^m$ which is defined as
\eq{
n_h = 
%(-1)^s 
\frac{1}{m!} \eps^{a_1 \dots a_m} \, T_{a_1} \w \dots \w T_{a_m}
}
where
%$s$ is the number of minuses in the signature $(r,s)$ of $\eta$ and 
$\{ T_a \}_{a = 1, \dots, m}$ is the standard $\eta$-orthonormal basis of $\R^m$.

Then we have two families of Hodge star operators
\eq{
&\map{\hod_k}{\Omega^k (M)}{\Omega^{m-k} (M)}{\pphi}{\hod_k \pphi}
	\\
&\map{\st_h}{\Gamma(\Lambda^h E)}{\Gamma(\Lambda^{m-h}E)}{v}{\st_h v}
}
which are uniquely defined by
\eq{
&\pphi \w \hod_k \psi = g(\pphi, \psi) \nu_g
	\\
&v \w \st_h w = h(v,w) n_h	
}
Notice that since $g$ has signature $(r,s)$ we have that $\hod_k$ is a \emph{signed} isometry, that is
\eq{
g(\hod_k \pphi, \hod_k \psi) = (-1)^{s} g(\pphi, \psi)
}
The factor $(-1)^{s}$ (i.e.\ the product of all minuses in the diagonal form of $g$) is often called the \emph{sign of $g$}. As a consequence of this fact we have the identities
\eq{
\hod_{m-k} (\hod_k \pphi) = (-1)^{s}(-1)^{k(m-k)} \id
}
so that 
\eq{
(\hod_k)^{-1} = (-1)^{s} (-1)^{k(m-k)} \hod_{m-k}
}
Another difference with respect to the Euclidean case is that
\eq{
\hod \nu_g = g(\nu_g, \nu_g) = (-1)^s
}

Identical considerations apply to $\st_h$ by susbstituting $\nu_g$ for $n_h$. See \cite{cb_dwm_1}, p.\, 294, for an exhaustive treatment on the subject.

The $\st$ operator and the representations $\Lambda^h T\ell$ of $\spin(r,s)$ on $\Lambda^h \R^m$ have an interesting interaction

\begin{prop}
\label{vec_forms_st_equiv}
On $\Lambda^h \R^m$ we have that $\st$ is an equivariant map for the representation $\Lambda^h T\ell \colon \spin(r,s) \ract \Lambda^h \R^m$.
\begin{proof}
First consider $S \in \Spin_0(r,s)$ and $v \in \Lambda^h \R^m$. We use the shorthand notation
\eq{
S \. v = \Lambda^h \ell(S)(v)
}
Then $\st \( S \. v \)$ is the element such that for any $w \in \Lambda^{m-h} \R^m$
\eq{
w \w \st(S \. v) 
&= q(w, S \.v )n_\eta
	\\
&= q(\inv S \. w, v)n_\eta 
	\\
&= (\inv S \. w) \w \st v
	\\
&= \inv S \. (w \w \st v) - w \w (\inv S \. \st v)
}
or
\eq{
w \w \[ \st (S \. v) + \inv S \. \st v \] = \inv S \. (w \w \st v)
}
Notice that 
\eq{
\inv S \. (w \w \st v) = q(w,v) \, \inv S \. n_\eta
}
Since $n_\eta$ is the volume form induced by $\eta$ we have 
\eq{
\inv S \. n_\eta = \det(\ell(\inv S)) n_\eta = n_\eta
}
And we have
\eq{
w \w \[ \st (S \. v) + \inv S \. \st v \] = q(w,v) n_\eta
}

We now consider the curve $\Gamma \colon \R \rarr \Spin_0(r,s)$ with $\Gamma(0) = 1$ and $\dot \Gamma(0) = X \in \spin(r,s)$. Since 
\eq{
X \. v = \Lambda^h T\ell(X)(v) = \left. \der{}{s} \Gamma(s) \. (v) \right.|_{s = 0}
}
we get
\eq{
&w \w \[ \st (X \. v) - X \. \st v \] = 0
	\\
&\rimpl w \w \st (X \. v) = w \w (X \. \st v)
}
Since we assumed an arbitrary $w \in \Lambda^h \R^m$ we have the thesis
\eq{
\st (X \. v) = X \. \st v
}

\end{proof}
\end{prop}

Using all of the above we now define a \emph{composite Hodge star operator} on $\Omega_H(Q, \ell)$
\begin{defn}[Composite Hodge Star]
Consider a decomposable element $\Phi = \pphi \ot v \in \Omega^{k,h}_H(Q, \ell)$, then the composite Hodge star operator $\bar \hod_{k,h}$ acts as
\eq{
\map{\bar \hod_{k,h}}{\Omega^{k,h}_H(Q, \ell)}{\Omega^{m-k, m-h}_H(Q, \ell)}{\Phi = \pphi \ot v}{\bar \hod_{k,h} \Phi = \hod_k \pphi \ot \st_h v}
}
\end{defn}
We can alternatively first define a metric $\bk{\.}{\.}$ on all of the $\Omega^{k,h}_H(Q, \ell)$ by prescribing it on decomposable elements as
\eq{
\bk{\pphi \ot v}{\psi \ot w} = g(\pphi, \psi) h(v, w)
}
and then we would have that the composite Hodge star operator satisfies
\eq{
\Phi \ow \bar \hod_{k,h} \Psi = \bk{\Phi}{\Psi} \nu_g \ot n_h = \bk{\Phi}{\Psi} \nu_e
}
where we defined the \emph{composite volume form} $\nu_e = \nu_g \ot n_h$. Notice how we can also define the action of the \q{partial} Hodge operators $\hod$ and $\st$ on $\Omega^{k,h}_H(Q, \ell)$, on decomposable elements these act as
\eq{
&\hod_k (\pphi \ot v) = (\hod_k \pphi) \ot v
	\\
&\st_h (\pphi \ot v) = \pphi \ot (\st_h v)
}
They also satisfy $\bar \hod_{k,h} = \st_h \circ \hod_k = \hod_k \circ \st_h$.

We finally state the properties of the composite Hodge star, which descend from the properties of $\hod$ and $\st$.

\begin{prop}[Composite Hodge Star]
\label{vec_forms_hodge_prop}
Let us consider $\Phi, \Psi \in \Omega^{k,h}_H(Q, \ell)$, then the composite Hodge star operator $\bar \hod_{k,h}$ satisfies
\begin{enumerate}[(i)]
\item $\Phi \ow \hod_{k} \Psi = \hod_{k} \Phi \ow \Psi$;

\item $\Phi \ow \st_h \Psi = \st_h \Phi \ow \Psi$;

\item $\Phi \ow \bar \hod_{k,h} \Psi = \bar \hod_{k,h} \Phi \ow \Psi$;

\item $(\bar \hod_{k,h})^{-1} = (-1)^{k(m-k)}(-1)^{h(m-h)} \bar \hod_{m-k, m-h}$;

\item $\bk{\.}{\.}$ is non degenerate;

\item $\bk{\bar \hod_{k,h} \Phi}{\bar \hod_{k,h} \Psi} = \bk{\Phi}{\Psi}$;

\item for any spin connection $\omega$ on $Q$ then $\D\omega (\st \Phi) = \st (\D\omega \Phi)$.
\end{enumerate}
\begin{proof}
For items $(\rom{1})-(\rom 6)$ it suffices to consider decomposable forms $\Phi = \pphi \ot v$ and $\Psi = \psi \ot w$ and use the corresponding properties of $\hod_k$ and $\st_h$. For the last item $(\rom 7)$, consider again a decomposable $\Phi = \pphi \ot v$. Then we have
\eq{
\D\omega(\st \Phi)
&= \D\omega(\pphi \ot \st v)
	\\
&= d(\pphi \ot \st v) + \omega \cw (\pphi \ot \st v)
	\\
&= d \pphi \ot \st v + \pphi \ot (\omega \cw \st v)
}
Using \cref{vec_forms_st_equiv} on the second term we get
\eq{
\D\omega(\st \Phi)
&= d \pphi \ot \st v + \pphi \ot \st(\omega \cw v)
	\\
&= \st \( d \pphi \ot v + \pphi \ot (\omega \cw v) \)
	\\
&= \st ( \D\omega \Phi )
}

\end{proof}
\end{prop}

Of particular interest is the action of $\bar \hod$ on the $k$-powers of the solder form $\theta$
\eq{
\theta^k = \theta^{\ow k} = \underbrace{\theta \ow \dots \ow \theta}_{k \text{ times}}
}
We begin by studying their norms.

\begin{lemma}[Square-norm of $\theta^k$]
We have
\eq{
\bk{\theta^k}{\theta^k} 
&= \frac{m! \, k!}{(m-k)!}
}
\begin{proof}
Denote by $\{ T_a \}$ the $\eta$-orthonormal basis of $\R^m$, then we can decompose the solder form $\theta \colon TQ \rarr \R^m$ as $\theta = \theta^a \ot T_a$ with $\theta^a \in \Omega^1_H(Q)$. By definition of $h$ on $\Omega^{1,1}_H(Q, \ell)$ we then have
\eq{
g(\theta^a, \theta^b)
&= \eta^{ab}	
}
Then we can compute $\bk{\theta^k}{\theta^k}$:
\eq{
\bk{\theta^k}{\theta^k}
&= \bk{(\theta^{a_1} \w \dots \w \theta^{a_k}) \ot (T_{a_1} \w \dots \w T_{a_k})}{(\theta^{b_1} \w \dots \w \theta^{b_k}) \ot (T_{b_1} \w \dots \w T_{b_k})}
}
for fixed indices $\{ a_i \}$ and $\{ b_j \}$ the scalar product above is
\eq{
&\det \eta^{a_i b_j} \. \det \eta_{a_i b_j}
}
In the $\eta$-orthonormal basis $\{T_a\}$, the matrix entries $\eta_{ab}$ and $\eta^{ab}$ are numerically equal. Therefore the product above is just $(\det \eta_{a_i b_j})^2$, which is the square of a minor of $\eta$. All minors of $\eta$ are $\pm 1$ or $0$, hence we only need to count the non null ones. Given that $\eta$ is diagonal, the non null minors are those for which the corresponding submatrix is still diagonal, that is for which the sets $\{ a_i \}$ and $\{ b_j \}$ coincide. Given we are summing with repetition we have that for any fixed $k$-uple of $\{ a_i \}$ there are $k!$ terms, and since the possible $\{ a_i \}$ are $\bin{m}{k} k!$ we finally get
\eq{
\bk{\theta^k}{\theta^k} 
&= \binom{m}{k} k! \, k!
	\\
&= \frac{m! \, k!}{(m-k)!}
}

\end{proof}
\end{lemma}

Notice that the composite volume form $\nu_e$ is a generator of $\Omega^{m,m}_H(Q, \ell)$, and so is the $m$-th power of the solder form $\theta^m$. Therefore we have that $\nu_e \propto \theta^m$ and one can actually prove the following:

\begin{prop}
The composite volume form $\nu_e$ is expressible in terms of the solder form $\theta$ as
\eq{
\nu_e = \frac{1}{m!} \theta^m
}
\begin{proof}
By definition we have
\eq{
\nu_e = \nu_g \ot n_h
}
We have the norms
\eq{
\bk{\nu_e}{\nu_e} 
&= g(\nu_g, \nu_g) h(n_h, n_h) 
	\\
&= (-1)^s (-1)^s
	\\
&= 1
}
And, by the previous lemma, $\bk{\theta^m}{\theta^m} = (m!)^2$. Since $\Omega^{m,m}_H(Q, \ell)$ is $1$-dimensional we must have $\theta^m = \pm m! \, \nu_e$ but since $\theta \colon TM \rarr E$ is an isomorphism of oriented vector bundles, for any positive oriented, $g$-orthonormal frame $e(q), q \in Q_x$, we have
\eq{
&\theta^m (e(q)) = \theta(e(q)_1) \w \dots \w \theta(e(q)_m) \text{ is a positively oriented frame in $E$}
}
On the other hand
\eq{
&\nu_e(e(q)) = \nu_g(e(q)) \ot n_h = n_h
}
Since the volume $m$-vector $n_h$ is positively oriented by definition, we have to choose the plus sign and get $\theta^m = m! \, \nu_e$.

\end{proof}
\end{prop}

We can use the characterization $\theta^m = m! \, \nu_e$ to compute the Hodge duals of the various powers $\theta^k$
\begin{prop}[Composite Hodge Dual of $\theta^k$]
We have that 
\eq{
\bar \hod \theta^k = \frac{k!}{(m-k)!} \theta^{m-k}
}
\begin{proof}
Using that
\eq{
\bk{\theta^k}{\theta^k} = \frac{m! k!}{(m-k)!}
}
we can then compute the Hodge duals, since
\eq{
\theta^k \ow \bar\hod(\theta^k)
&= \bk{\theta^k}{\theta^k} \, \nu_e  
	\\
&= \frac{\cancel{m!} \, k!}{(m-k)!} \, \frac{1}{\cancel{m!}} \theta^m
	\\
&= \frac{k!}{(m-k)!} \theta^k \ow \theta^{m-k}
}
This implies, by the fact that Hodge duality is an isomorphism, that 
\eq{
\bar \hod (\theta^k) = \frac{k!}{(m-k)!} \theta^{m-k}
}
\end{proof}
\end{prop}

%\red{
%\eq{
%\bk{\theta^{m-k}}{\theta^{m-k}} 
%&= \frac{m! \, (m-k)!}{k!}
%	\\
%&= \frac{(m-k)!}{k!} \frac{(m-k)!}{k!} \bk{\theta^k}{\theta^k}
%	\\
%&= \frac{(m-k)!}{k!} \frac{\cancel{(m-k)!}}{\bcancel{k!}} \frac{m! \, \bcancel{k!}}{\cancel{(m-k)!}}
%}
%ok, torna tutto
%}

\begin{cor}
If $m = 2n$ is even, the power $\theta^n$ is selfdual with respect to $\bar \hod$. It also satisfies $\hod \theta^n = (-1)^s (-1)^{n^2} \st \theta^n$.
\begin{proof}
For the first part we just need to write $\bar \hod \theta^k$ for $m = 2n$ and $k = n$. For the second part, we have that $\bar \hod_{n,n} = \st_n \circ \hod_n$ so that
\eq{
\bar \hod \theta^n = \theta^n \rimpl \hod_n \theta^n = \inv\st_n (\theta^n)
}
The last identity can be simplified by expanding $(\st_n)^{-1}$ which is
\eq{
\hod_n \theta^n 
&= (-1)^{s} (-1)^{n(m-n)} \st \theta^n
	\\
&= (-1)^{s} (-1)^{n^2} \st \theta^n	
}
which is the thesis.
\end{proof}
\end{cor}
\begin{rem}
In the case of interest for Loop Quantum Gravity we have $m = 4$ (so that $n = 2$) and $(r,s) = (3,1)$ gives the sign of $g$ as $(-1)^s = -1$. Therefore we are left with
\eq{
\hod \theta^2 = - \st \theta^2
}
\end{rem}

\subsection{Local Expressions for Hodge Duals}

To rewrite the Holst Lagrangian in intrinsic form we will need to work in reverse from its usual local expression. To this end we recall the local expressions involved in Hodge duals, starting from the volume forms
\eq{
&\nu_g = \frac{\sqrt g}{m!} \epsilon_{\mu_1 \dots \mu_m} \, dx^{\mu_1} \w \dots \w dx^{\mu_m}
	\\
&n_h = \frac{1}{m!} \epsilon^{a_1 \dots a_m} \, T_{a_1} \w \dots \w T_{a_m}	
}
As is usual, $\sqrt g$ is a shorthand notation for $\sqrt{\abs{\det g}}$ and $\epsilon^{a_1 \dots a_m}$ are the coefficients of the totally contravariant Levi--Civita tensor density on $\R^m$.

A $k$-form $\pphi$ on $M$ and an $h$-vector $v$ in $\Lambda^h E$ are expressed locally as
\eq{
&\pphi = \frac{1}{k!} \pphi_{\mu_1 \dots \mu_k} \, dx^{\mu_1} \w \dots \w dx^{\mu_k}
	\\
&v = \frac{1}{h!} v^{a_1 \dots a_h} \, T_{a_1} \w \dots \w T_{a_h}	
}
The action of the star operator $\hod_k$ is
\eq{
(\hod \pphi)_{\mu_{k+1} \dots \mu_m} = \frac{\sqrt g}{k!} \epsilon^{\mu_1}_\.{}^{\dots}_\.{}^{\mu_k}_\.{}_{\mu_{k+1} \dots \mu_m} \, \pphi_{\mu_1 \dots \mu_k}
}
where dots denote greek indices raised/lowered through the metric $g$, i.e.\
\eq{
\epsilon^{\mu_1}_\.{}^{\dots}_\.{}^{\mu_k}_\.{}_{\mu_{k+1} \dots \mu_m} = g^{\mu_1 \alpha_1} \dots g^{\mu_k \alpha_k} \epsilon_{\alpha_1 \dots \alpha_k \mu_{k+1} \dots \mu_m}
}
Therefore
\eq{
\hod \pphi = \frac{1}{(m-k)!} (\hod \pphi)_{\mu_{k+1} \dots \mu_m} \, dx^{\mu_{k+1}} \w \dots \w dx^{\mu_m}
}

Similarly, for $\st_h$, we have
\eq{
(\st v)^{a_{h+1} \dots a_m} = 
%(-1)^s 
\frac{1}{h!} \epsilon_{a_1}^\.{}_{\dots}^\.{}_{a_h}^\.{}^{a_{h+1} \dots a_m} \, v^{a_1 \dots a_h}
}
where dots denote latin indices raised/lowered through the metric $\eta$, therefore
\eq{
\st v = \frac{1}{(m-h)!} (\st v)^{a_{h+1} \dots a_m} \, T_{a_{h+1}} \w \dots \w T_{a_m}
}

%%%% %%%
%\subsection{The case $m = 4$ and $(r,s) = (3,1)$}\label[ssec]{vec_forms_4_dim_lor}
%
%We now specialize the discourse to the case of interest for LQG: a $4$-dimensional manifold $M$ which is lorentzian, i.e.\ it admits metrics of signature $(3,1)$.
%
%Since the sign of $g$ is $(-1)^1 = -1$ we have that $\hod$ is an anti-isometry
%\eq{
%g(\hod \pphi, \hod \psi) = -g(\pphi, \psi), \quad \forall \pphi, \psi \in \Omega^k(M)
%}
%The inverse Hodge operators are given by
%\eq{
%(\hod_k)^{-1} 
%&= -(-1)^{k(4-k)} \hod_{4-k}
%	\\
%&= (-1)^{k^2 + 1} \hod_{4-k}	
%}
%In particular, the volume form $\nu_g$ has negative unitary norm
%\eq{
%g(\nu_g , \nu_g) = -1
%}
%Analoguous considerations are true for $\st$.
%
%Manifold dimension $4$ is peculiar, since on $\Lambda^2 \R^4 \simeq \spin(3,1)$ the operator $\st$ is an isomorphism. In particular one can study the relation between $\st$ and $\ad$
%\begin{prop}
%On $\spin(3,1)$ we have that $\st \circ \ad_X = \ad_X \circ \, \st$ for any $X \in \spin(3,1)$.
%\begin{proof}
%Consider $X, Y, Z \in \spin(3,1)$, then
%\eq{
%\st (\ad_X Y) 
%&= \st [X, Y]
%}
%is the element such that
%\eq{
%Z \w \st[X, Y] 
%&= q(Z, [X, Y]) n_\eta
%	\\
%&= q([Z, X], Y) n_\eta
%}
%On the other hand
%\eq{
%Z \w (\ad_X \st Y)
%&= \ad_X(Z \w \st Y) - (\ad_X Z) \w \st Y
%}
%The first term on the r.h.s.\ is
%\eq{
%-q(Z, Y) \ad_X(T_0 \w T_1 \w T_2 \w T_3)
%&= -q(Z,Y)
%}
%
%\end{proof}
%\end{prop}

%
\section{The Trace of a Form in $\Omega^{k,h}_H(Q, \ell)$}\label[sec]{vec_forms_trace_sec}

Consider $\Phi \in \Omega^{k,h}_H(Q, \ell)$ with local coordinates $\Phi^{a_1 \dots a_h}_{\mu_1 \dots \mu_k}$, that is 
\eq{
\Phi = \frac{1}{k! \, h!} \Phi^{a_1 \dots a_h}_{\mu_1 \dots \mu_k} \, (dx^{\mu_1} \w \dots dx^{\mu_k}) \ot (T_{a_1} \w \dots T_{a_h})
}
We define the \emph{trace of $\Phi$} as the form $\tr \Phi$ which is $0$ whenever $k = 0$ or $h = 0$ and otherwise it is the form in $\Omega^{k-1,h-1}_H(Q, \ell)$ with coordinates
\eq{
(\tr \Phi)^{a_1 \dots a_{h-1}}_{\mu_1 \dots \mu_{k-1}} = \Phi^{a a_1 \dots a_{h-1}}_{\mu \mu_1 \dots \mu_{k-1}} \, e^\mu_a
}
This can be formulated in intrinsic language using $\ow$ and $\bar \hod$. We can, however, abstract the situation slightly: for any element $f \in \lie X(M) \ot \Gamma(E^*)$ we can define the interior product $- \chair f$:

\begin{defn}[Interior Product by $f$]
Define $\Omega^{k,h}_H(Q, \ell) = 0$ whenever $k < 0$ or $h < 0$. For any $f \in \lie X(M) \ot \Gamma(E^*)$ we have the interior product
\eq{
\map{- \chair f}{\Omega^{k,h}(Q, \ell)}{\Omega^{k-1, h-1}(Q, \ell)}{\Phi}{\Phi \chair f}
}
which is defined on decomposable $\Phi = \pphi \ot v$ and $f = \xi \ot \alpha$ as
\eq{
\Phi \chair f =  (\pphi \chair \xi) \ot (v \chair \alpha)
}
\end{defn}

%In local coordinates we have 
%\eq{
%\Phi \chair f 
%&= \frac{1}{k! \, h!} \Phi^{a a_1 \dots a_{h-1}}_{\mu \mu_1 \dots \mu_{k-1}} \, f^\nu_b  \[ (dx^\mu \w dx^{\mu_1} \w \dots \w dx^{\mu_{k-1}}) \chair \d_\nu \] \ot \[ (T_a \w T_{a_1} \w \dots \w T_{a_{k-1}} ) \chair T_b \]
%	\\
%&= \frac{1}{k \. h} \Phi^{a a_1 \dots a_{h-1}}_{\mu \mu_1 \dots \mu_{k-1}} \, f^\mu_a  \( dx^{\mu_1} \w \dots \w dx^{\mu_{k-1}} \) \ot \( T_a \w T_{a_1} \w \dots \w T_{a_{k-1}} \)
%	\\
%&= \frac{1}{(k-1)! (h-1)!} \( k! \, h!  \, \Phi^{a a_1 \dots a_{h-1}}_{\mu \mu_1 \dots \mu_{k-1}} \, f^\mu_a \)  \( dx^{\mu_1} \w \dots \w dx^{\mu_{k-1}} \) \ot \( T_a \w T_{a_1} \w \dots \w T_{a_{k-1}} \)	
%}

Now, the spaces $\Omega^{1,1}_H(Q, \ell)$ and $\lie X(M) \ot \Gamma(E^*)$ are dual to each other, therefore the scalar product $\bk{-}{-}$ induces, as usual, the sharp isomorphism $\sharp$
\eq{
\map{\sharp}{\Omega^{1,1}_H(Q, \ell)}{\lie X(M) \ot \Gamma(E^*)}{\Phi}{\Phi^\sharp = \bk{\Phi}{-}}
}
and also the flat isomorphism $\flat = \inv\sharp$. Notice that if $\Phi = \pphi \ot v$ then
\eq{
\Phi^\sharp = \pphi^{\sharp_g} \ot v^{\flat_h}
}
where the $g$ subscript denotes the $\sharp$ and $\flat$ isomorphisms in $\Omega^k_H(Q)$ induced by $g$. Similarly for the subscript $h$.

%, since $\bk{\theta}{\theta} = m$ and
%\eq{
%e(\theta) 
%&= e^\mu_a \, \theta^a_\mu 
%	\\
%&= e^\mu_a \, \eta^{ab} e_b^\nu g_{\mu\nu}
%	\\
%&= g^{\mu\nu} g_{\mu\nu}	
%	\\
%&= m
%}
%Therefore $e(\theta) = \bk{\theta}{\theta}$, which is the definition of $\theta^\sharp$

We can now state and prove the following property, which shows the interplay between the interior product, the musical isomorphisms, and the composite Hodge dual.

\begin{prop}[Intrinsic Definition of Interior Product]
We have that
\eq{
\Phi \chair f
&= (\bar \hod)^{-1} (\bar \hod \Phi \ow f^\flat)
}
%which implies
%\eq{
%\tr \Phi 
%&= (-1)^{k+h}(-1)^{(m-k+1)(k-1)}(-1)^{(m-h+1)(h-1)} \, \bar \hod (\theta \ow \bar \hod \Phi)
%	\\
%&= (-1)^{m(k+h-2) - (k-1)^2 - (h-1)^2 + k + h} \, \bar \hod (\theta \ow \bar \hod \Phi)
%	\\
%&= (-1)^{m(k+h) - k^2 + 3 k - h^2 + 3 h} \, \bar \hod (\theta \ow \bar \hod \Phi)	
%	\\
%&= (-1)^{(m+3)(k+h) - (k + h)^2 + 2kh} \, \bar \hod (\theta \ow \bar \hod \Phi)	
%	\\
%&= (-1)^{(m-k-h+1)(k+h)} \, \bar \hod (\theta \ow \bar \hod \Phi)	
%}
\begin{proof}
As usual, it suffices to consider decomposable elements $\Phi = \pphi \ot v$ and $f = \xi \ot \alpha$. We will use the following property of Hodge duals: consider $\pphi \in \Omega^k(M)$ and $\beta \in \Omega^1(M)$, then we have
\eq{
\hod(\pphi \w \beta) =  \hod \pphi \chair \beta^{\sharp_g}
}
Similarly consider $v \in \Lambda^h E$ and $x \in \Lambda^1 E = E$, we have
\eq{
\st(v \w x) = \st v \chair x^{\flat_h}
}
which, by definition, gives
\eq{
\bar \hod \[ (\pphi \ot v) \ow (\beta \ot x)^\sharp \] = \bar \hod (\pphi \ot v) \chair (\beta \ot x)
}

If $k = 0$ or $h = 0$ then there is nothing to prove, otherwise
\eq{
\Phi \chair f
&= (\pphi \chair \xi) \ot (v \chair \alpha)
	\\
&= (-1)^{k(m-k)}(\hod (\hod \pphi) \chair \xi) \ot (-1)^{h (m-h)}(\st (\st v) \chair \alpha)
	\\
&= (-1)^{k(m-k)}\hod (\hod \pphi \w \xi^{\flat_g} ) \ot (-1)^{h (m-h)}\st (\st v \w \alpha^{\sharp_h})
	\\
&= (-1)^{k(m-k)}(-1)^{h (m-h)} \bar \hod \[ (\hod \pphi \ot \st v) \ow (\xi^{\flat_g} \ot \alpha^{\sharp_h}) \]
	\\
&= \inv{\bar \hod} (\bar \hod \Phi \ow f^\flat )	
}
which is the thesis.

\end{proof}
\end{prop}

\begin{cor}
\label{vec_forms_trace}
The intrinsic definition of the trace is
\eq{
\tr \Phi = \Phi \chair \theta= \inv{\bar \hod}(\bar \hod \Phi \ow \theta)
}
\end{cor}

%\begin{prop}[Interior Product]
%The interior product $f \chair -$ satisfies:
%\eq{
%f \chair (\Phi \ow \Psi) = \red{???}
%}
%\begin{proof}
%This follows from the analogous statement for \q{classical} forms. Again we will prove the proposition for decomposable elements, $\Phi = \pphi \ot v$, $\Psi = \psi \ot w$, and $f = \xi \ot \alpha$. We have
%\eq{
%f \chair (\Phi \ow \Psi)
%&= [\xi \chair (\pphi \w \psi) ] \ot [\alpha \chair (v \w w)]
%	\\
%&= [(\xi \chair \pphi ) \w \psi) + (-1)^k \pphi \w (\xi \chair \psi)] \ot [(\alpha \chair v) \w w + (-1)^h v \w (\alpha \chair w)]	
%	\\
%&= (f \chair \Phi) \ow \Psi + (-1)^{k+h} \, \Phi \ow (f \chair \Psi) + 
%	\\
%&\quad + (-1)^h [(\xi \chair \pphi) \w \psi] \ot [v \w (\alpha \chair w)] + (-1)^k [\pphi \w (\xi \chair \psi)] \ot [(\alpha \chair v) \w w]
%	\\
%&= (f \chair \Phi) \ow \Psi + (-1)^{k+h} \, \Phi \ow (f \chair \Psi) + 
%	\\
%&\quad + (-1)^h [\inv \hod (\hod \pphi \w \xi^{\flat_g}) \w \psi] \ot [v \w \inv \st (\st w \w \alpha^{\sharp_h})] +
%	\\
%&\quad + (-1)^k [\pphi \w (\xi \chair \psi)] \ot [(\alpha \chair v) \w w]	
%	\\
%&= (f \chair \Phi) \ow \Psi + (-1)^{k+h} \, \Phi \ow (f \chair \Psi) + 
%	\\
%&\quad + (-1)^h (-1)^{(m-k+1)(k-1)} \hod [\psi^{\sharp_g} \chair (\hod \pphi \w \xi^{\flat_g})] \ot [v \w \inv \st (\st w \w \alpha^{\sharp_h})] +
%	\\
%&\quad + (-1)^k [\pphi \w (\xi \chair \psi)] \ot [(\alpha \chair v) \w w]		
%}
%\red{
%da fare come sotto?
%}
%\end{proof}
%\end{prop}

One immediate question is the following: is $\theta \ow \tr \Phi$ proportional to $\Phi$? That is, is $\theta \ow -$ proportional to an inverse of the trace? The answer in general is negative, as we now prove.

\begin{lemma}[Trace Lemma]
For any $\Phi \in \Omega^{k,h}_H(Q, \ell)$ we have the following identity
\eq{
\tr (\theta \ow \Phi) = \theta \ow \tr \Phi  + (m - k - h) \Phi
}
\begin{proof}
The rather lenghty proof can be found in \cref{app_trace_lemma}.

\end{proof}
\end{lemma}

\begin{prop}[Injectivity of $\theta \ow -$]
\label{vec_forms_owtheta_inj}
For $k + h < m$ the map
\eq{
\map{\theta \ow -}{\Omega^{k,h}_H(Q, \ell)}{\Omega^{k+1,h+1}_H(Q, \ell)}{\Phi}{\theta \ow \Phi}
}
is injective.
\begin{proof}
We first compute iterated traces by induction
\eq{
\tr^{r+1}(\theta \ow \Phi) = \theta \ow \tr^{r+1}\Phi + (r+1)(m - k - h + r) \tr^r \Phi
}
The base case $r = 0$ is the previous lemma, then
\eq{
\tr^{r+1}(\theta \ow \Phi) 
&= \tr (\tr^{r+1} (\theta \ow \Phi))
	\\
&= \tr (\theta \ow \tr^{r}\Phi + (r)(m - k - h + r - 1) \tr^{r-1} \Phi)
	\\
&= \theta \ow \tr^{r+1}\Phi + (m - (k - r) - (h - r)) \tr^r \Phi + (r)(m - k - h + r - 1) \tr^{r} \Phi		
	\\
&= \theta \ow \tr^{r+1}\Phi + (r + 1)(m - k - h) \tr^r \Phi + (2r + r^2 - r) \tr^{r} \Phi		
	\\
&= \theta \ow \tr^{r+1}\Phi + (r + 1)(m - k - h + r) \tr^r \Phi
}
and the induction is complete.

Now suppose $\theta \ow \Phi = 0$, the identities above reduce to
\eq{
(r+1)(m - k - h + r) \tr^r \Phi = -\theta \ow \tr^{r+1}\Phi
}
Recall that for $r > \min \{ k,h \}$ the traces are, by definition, all zero. Denote by $p = m-k-h$ and by $q = \min \{ k,h \}$, then we have the tower of identities
\eq{
&p \Phi = -\theta \ow \tr\Phi
	\\
&2(p + 1) \tr \Phi = -\theta \ow \tr^2\Phi
	\\
&\vdots
	\\
&(q+1)(p + q) \tr^q \Phi = 0
}
If $p + q = 0$ then either $k = m$ or $h = m$, which implies $h < 0$ or $k < 0$ (since $k + h < m$) therefore $\theta \ow \Phi$ is always zero and there is nothing to prove. Otherwise $p + q > 0$ (we are using that $k + h < m \iff p > 0$) and we can solve the system of equation backwards
\eq{
\tr^q \Phi = \tr^{q-1} \Phi = \dots = \tr \Phi = \Phi = 0
}
which is the thesis.

\end{proof}
\end{prop}

\begin{rem}
We actually proved a stronger result, that is: for $\Phi \in \Omega^{k,h}_H(Q, \ell)$ with $k + h < m$ and $q = \min\{ k,h \}$ we have that
\eq{
\theta \ow \Phi = 0 \rimpl \tr^i \Phi = 0, \quad \forall i = 0, \dots, q
}

Also, in the case $k + h = m \iff p = 0$ we have the partial result
\eq{
\theta \ow \Phi = 0 \rimpl \tr^i \Phi = 0, \quad \forall i = 1, \dots, q
}
\end{rem}

Similarly one can prove

\begin{prop}[Injectivity of the Trace]
\label{vec_forms_trace_inj}
For $k + h > m$ the map
\eq{
\map{\tr}{\Omega^{k,h}_H(Q, \ell)}{\Omega^{k-1,h-1}_H(Q, \ell)}{\Phi}{\tr \Phi}
}
is injective. In particular, for $q = \min\{ m-k, m-h \}$ we have
\eq{
\tr \Phi = 0 \rimpl \theta^i \ow \Phi = 0, \quad \forall i = 0, \dots, q
}
while for $k + h = m \iff p = m - k - h = 0$ we have
\eq{
\tr \Phi = 0 \rimpl \theta^i \ow \Phi = 0, \quad \forall i = 1, \dots, q
}
\begin{proof}
This proof is similar to that of the property above and is proved \cref{app_trace_lemma}.
\end{proof}
\end{prop}

%
%% %%
\section{Recasting the Holst Lagrangian}\label[sec]{bi_dynamical_holst}

Throughout this chapter we fix a spin frame $e \colon Q \rarr L(M)$, where $\sv Q = (Q, p, M, \Spin_0(3,1))$ is a $\Spin_0(3,1)$-bundle on a connected, orientable, $4$-dimensional lorentzian manifold $M$. We denote by $\theta$ the solder form associated to the spin frame $e$, if $\sv E = Q^\ell = (E, p^\ell, M, \R^4)$ is the vector bundle associated to $Q$ via the action $\ell \colon \Spin_0(3,1) \rarr \GL(4)$ then $\theta \colon TM \rarr E$ or $\theta$ is a section of the bundle $T^*M \ot_E E$ over $M$.

Let us start from the usual expression in local coordinates and work out its intrinsic equivalent. The Holst Lagrangian with Holst parameter $\gamma \neq 0$ is (see \cite{ffr2})
\eq{
\sv L_\gamma(\theta, j^1 \omega) = \frac{1}{4\bar G} \[ \epsilon_{abcd} R^{ab} \w \theta^c \w \theta^d + \frac{2}{\gamma} R^{ab}{} \w \theta^\._{a} \w \theta^\._{b} \]
}
where
\begin{itemize}
\item $\bar G$ includes all physical constants and is truly meaningful only when coupling with matter;

\item $\epsilon_{abcd}$ is the $4$-dimensional Levi--Civita symbol in $\R^4$;

\item $\theta^a$ are the coefficients of the solder form $\theta = \theta^a \ot T_a$;

\item $R^{ab}$ are the coefficients of the curvature of a spin connection $\omega = \half 1\omega^{ab} \ot J_{ab}$ on $Q$. In particular
\eq{
R = \D\omega \omega 
&= \half 1 \D\omega \omega^{ab} \ot J_{ab}
	\\
&= \half 1 \( d\omega^{ab} + \half 1 \omega^{\.}_c{}^{[a} \omega^{\uline c b]} \) \ot J_{ab}
}
\end{itemize}
The Holst lagrangian then is of first-order in the spin connection $\omega$ and of order zero in the spin frame.

Working a little bit we have
\eq{
\sv L_\gamma
&= \frac{1}{4\bar G} \[ \epsilon_{abcd} R^{ab} \w \theta^c \w \theta^d + \frac{2}{\gamma} R^{ab} \w \eta_{ac} \theta^c \w \eta_{bd} \theta^d \] 
	\\
&= \frac{1}{4\bar G} \[ 2R^{ab} \eta_{ae} \eta_{bf} \w \( \half 1 \epsilon^{ef}_{\.\.}{}_{cd}  (\theta^2)^{cd} + \frac{1}{\gamma} (\theta^2)^{ef} \) \] 	
}
Using the complete skew-symmetry of the Levi--Civita symbol $\epsilon$ and the property
\eq{
\epsilon^{ef}_{\.\.}{}_{cd} = - \det \eta \, \epsilon^{ef}{}^{\.\.}_{cd}
}
we then get
\eq{
\sv L_\gamma
&= \frac{1}{4\bar G} \[ 2R^{ab} \eta_{e[a} \eta_{b]f} \w \( (\st \theta^2)^{ef} + \frac{1}{\gamma} (\theta^2)^{ef} \) \] 
	\\
&= \frac{1}{4\bar G} \[ 2R^{ab} \eta_{e[a} \eta_{b]f} \w \( -(\hod \theta^2)^{ef} + \frac{1}{\gamma} (\bar \hod\theta^2)^{ef} \) \] 
	\\
&= \frac{1}{4\bar G} \[ 2R^{ab} \eta_{e[a} \eta_{b]f} \w \hod \( -(\theta^2)^{ef} + \frac{1}{\gamma} (\st\theta^2)^{ef} \) \] 	
	\\
&= \frac{1}{4\bar G} \, g \(R^{ab} , -(\theta^2)^{ef} + \frac{1}{\gamma} (\st\theta^2)^{ef} \) \nu_g  \ot \eta_{e[a} \eta_{b]f}
}
where $g$ is the metric induced on $\Omega(Q)$ by the metric $g$ on $M$ induced by the solder form $\theta$.
Recall that
\eq{
2 \eta_{e[a} \eta_{b] f} 
&= \eta_{ae} \eta_{bf} - \eta_{af} \eta_{be} 
	\\
&= \begin{vmatrix}
\eta_{ae} & \eta_{af}
	\\
\eta_{be} & \eta_{bf}
\end{vmatrix} 
	\\
&= \eta(T_a \w T_b, T_e \w T_f)
	\\
&= -q(J_{ab}, J_{ef}) \, \st n_h
}
Therefore, we can write
\eq{
\sv L_\gamma 
&= -\frac{1}{4\bar G} \, g \(R^{ab} , -(\theta^2)^{ef} + \frac{1}{\gamma} (\st\theta^2)^{ef} \) \nu_g  \ot \eta(J_{ab}, J_{ef}) \, \st n_h
	\\
&= \st \[ \frac{1}{4\bar G} \, R \ow \( \bar \hod \theta^2 - \frac{1}{\gamma} \bar \hod (\st\theta^2)^{ef} \) \]
	\\
&= \st \[ \frac{1}{4\bar G} \, R \ow \( \theta^2 - \frac{1}{\gamma} (-\hod\theta^2)^{ef} \) \]
}
and finally
\eq{
\sv L_\gamma = \frac{1}{4\bar G} \, \st \[ R \ow \( 1 - \frac{1}{\gamma} \st \) \theta^2 \]
}

We will denote the operator $1 - \tfrac{1}{\gamma} \st$ by $\st_\gamma$, which can be defined for any dimension $m = \dim M$ and any signature. 

However, since $\st \colon \Omega^{k,h}_H(Q, \ell) \rarr \Omega^{k,m-h}(Q, \ell)$, we have that in general
\eq{
\st_\gamma \colon \Omega^{k,h}_H(Q, \ell) \rarr \Omega^{k,h}_H(Q, \ell) \oplus \Omega^{k,m-h}_H(Q, \ell)
}
For $m = 4$ and $h = 2$ the operator $\st_\gamma$ is an endomorphism of $\Omega^{k,2}_H(Q, \ell)$ and we can show that in the lorentzian case $(r, s) = (3,1)$ it is an isomorphism for any $\gamma \neq 0$.

\begin{prop}[$\st_\gamma$ Operator]
For $m = 4$ and $(r,s) = (3,1)$ the operator 
\eq{
\map{\st_\gamma}{\Omega^{k,2}_H(Q, \ell)}{\Omega^{k, 2}_H(Q, \ell)}{\Phi}{\st_\gamma \Phi = \( 1 - \frac{1}{\gamma} \st \) \Phi}
}
is an isomorphism.
\begin{proof}
If $\Psi = \st_\gamma \Phi$ then
\eq{
\st \Psi 
&= \( \st - \frac{1}{\gamma}(-1)^s (-1)^{n^2}\) \Phi
	\\
&= -\gamma\( 1 - 1 - \frac{1}{\gamma}\st \) \Phi - \frac{1}{\gamma}(-1)^s (-1)^{n^2} \Phi
	\\
&= -\gamma \Psi + \gamma \Phi + (-1)^{1 + s + n^2}\frac{1}{\gamma} \Phi
}
Thus
\eq{
&\( \gamma + (-1)^{1 + s + n^2}\frac{1}{\gamma} \) \Phi = (\gamma + \st)\Psi 
}
By specializing to $s = 1, n = 2$, the inverse to $\st_\gamma$ is
\eq{
(\st_\gamma)^{-1} = \frac{\gamma (\gamma + \st)}{\gamma^2 + 1}
}
\end{proof}
\end{prop}

\section{Holst Lagrangian in terms of BI Connections: The Barbero--Immirzi--Holst (BIH) Lagrangian}\label[sec]{bi_dynamical_bih}

As of now the Holst Lagrangian depends on the solder form $\theta$ to order $0$, and on the spin connection $\omega$ to order $1$. The main scope of this chapter is to recast the Holst Lagrangian into new fields by splitting the spin connection $\omega$ into a pair $(A, \kappa)$ where $A$ is the BI connection and $\kappa$ is the extrinsic spacetime field. 
%Complete characterization of this splitting was given in \cref{orizz_bi_conn}, we recall it briefly.

As in \cref{spin_su_red_pair}, if we fix an Immirzi parameter $\beta \in \R$ we can decompose the spin connection $\omega = \half 1\omega^{ab} \ot J_{ab}$ into its $\su(2)$-part $A$ and its $\lie m_\beta$-part $\kappa$, which in components are
\eq{
\sys{
A = A^k \, L_k
	\\
\kappa = \kappa^k \, H^{(\beta)}_k	
} \iff \sys{
A^k = \half 1 \epsilon_{ij}{}^k \, \omega^{ij} + \beta \omega^{0k}
	\\
\kappa^k = \omega^{0k}
}
}
where the generators satisfy the following commutation relations
\eq{
&[L_i, L_j] = \epsilon_{ij}{}^k \, L_k
	\\
&[L_i, H^{(\beta)}_j] = \epsilon_{ij}{}^k \, H^{(\beta)}_k
	\\
&[H^{(\beta)}_i, H^{(\beta)}_j] = -\epsilon_{ij}{}^k \[ (1 + \beta^2) L_k + 2\beta H^{(\beta)}_k \]
}

Since any spin frame $(Q, e)$ on $M$ always admits a reduction to a $\SU(2)$-bundle ${}^+ Q$ on $M$, by restriction, $A$ is an $\SU(2)$-connection on ${}^+ Q$ and $\kappa$ is a tensorial $1$-form of type $(\Ad_{\Spin(3,1)}(\SU(2)), \lie m_\beta)$, the adjoint action of $\Spin(3,1)$ restricted to $\SU(2)$ on $\lie m_\beta$. We denote by ${}^+ E^\beta$ the vector bundle ${}^+ Q \times_{\Ad} \lie m_\beta$ associated to ${}^+ Q$ via the representation $\Ad \colon \Spin(3,1) \rarr \GL(\lie m_\beta)$ restricted to $\SU(2)$.

Since $\omega = A + \kappa$, when we restrict the Holst lagrangian to ${}^+ Q$ it can be recast as
\eq{
\sv L_\gamma 
&= \frac{1}{4\bar G} \st \[ \(d\omega + \half 1[\omega \w \omega]\) \ow \st_\gamma \theta^2 \]
	\\
&= \frac{1}{4\bar G} \st \[ \(dA + dK + \half 1[(A + \kappa) \w (A + \kappa)]\) \ow \st_\gamma \theta^2 \]
	\\
&= \frac{1}{4\bar G} \st \[ \(dA + \half 1[A, A] + dK + \half 1 [A \w \kappa] + \half 1 [\kappa \w A] + \half 1 [\kappa \w \kappa]\)  \ow \st_\gamma \theta^2 \]
}

Denoting by $F = \D A A$ the field-strength/curvature of the BI connection and noticing that $[\kappa \w A] = [A \w \kappa]$ we get
\eq{
R = F + \D A \kappa + \half 1 [\kappa \w \kappa]
} 
so that we get the Barbero--Immirzi--Holst (BIH) lagrangian, which is the Holst lagrangian in the fields $(\theta, A, \kappa)$:
\eq{
\sv L_\gamma(\theta, j^1 A, j^1 \kappa)
&= \frac{1}{4\bar G} \st \[ \( F + \D A \kappa + \half 1 [\kappa \w \kappa]\)  \ow \st_\gamma \theta^2 \]
}

\section{Variational Analysis of the Barbero--Immirzi--Holst Lagrangian}\label[sec]{bi_dynamical_bih_el}

%To compute the variation of the Holst Lagrangian $\sv L_\gamma$ we will use many different lemmas. To not break the flow of this section, we merely state the lemmas when needed, postponing their proofs to the \cref{app_var_calc}.

As prescribed by the principle of general relativity, we have to vary with respect to all the fields, namely the solder form $\theta$, the BI connection $A$, and the extrinsic spacetime field $\kappa$. The solder form is a section of $T^*M \ot_M E$, or equivalently, an element of $\Omega^{1}_H(Q, \ell)$. The BI connection $A$ is a section of $\Con_{\SU(2)}({}^+ Q)$, the bundle of principal connections on the $\SU(2)$-bundle ${}^+ Q$, it is an affine bundle \emph{on $M$} (see \cite{fatifranca}, p.\ 94). The extrinsic spacetime field $\kappa$ is a section of $T^*M \ot_M {}^+ E^\beta$.

The configuration bundle of the BIH lagrangian $\sv L_\gamma$ then will be the product bundle
\eq{
\sv C = (T^*M \ot_M E) \times_M (\Con_{\SU(2)}({}^+ Q)) \times_M (T^*M \ot_M {}^+ E^\beta)
}
A variation then is a section $X \colon M \rarr V\sv C$ of the bundle of vertical vectors over $\sv C$, which is supported on a compact region $D \subset M$ and is zero on the boundary $\d D$. Since $\sv C$ is a product bundle we have
\eq{
V\sv C = V(T^*M \ot_M E) \oplus_M V(\Con_{\SU(2)}({}^+ Q)) \oplus_M V(T^*M \ot_M {}^+ E^\beta)
}
That is, any variation is the direct sum of three \q{basic} variations. We can prove that the variations of the three bundles are tensorial $1$-forms of the appropriate type.

\begin{lemma}
Consider an affine bundle $\sv B = (B, \pi, M, A)$ on $M$ which is modelled on a vector bundle $\sv E = (E, \pi', M, V)$, that is for each $B_x$ we have a map
\eq{
\mapp{B_x \times E_x}{B_x}{(a_x, v_x)}{a_x + v_x}
}
Then the vertical bundle $VB$ on $B$ is isomorphic to $B \times_M E$.
\begin{proof}
Using the affine structure of $B$ we can build a number of vertical curves $\gamma_x \colon (-\eps, \eps) \rarr B_x$ based in $a_x = \gamma_x(0)$ this way: for any $v_x \in E_x$ define
\eq{
\gamma(a_x, v_x)(s) = a_x + s \. v_x
}
Then for every smooth function $f \colon B \rarr \R$ the vector $[\gamma(a_x, v_x)]_{a_x}$ tangent to $a_x$ corresponding to the class of $\gamma(a_x, v_x)$ acts as
\eq{
[\gamma(a_x, v_x)]_{a_x} f = \left. \der{}{s} f(a_x + s \. v_x ) \right|_{s = 0}
}

The resulting tangent vector is zero iff
\eq{
\left. \der{}{s} f(a_x + s \. v_x ) \right|_{s = 0}
&= 0
	\\
&= \left. \der{}{s} f(a_x + s \. 0 ) \right|_{s = 0}
}

So that the correspondence between $B_x \times E_x$ and $V_{a_x} B$ given by
\eq{
(a_x, v_x) \rarrto [\gamma(a_x, v_x)]_{a_x}
}
is one-to-one. By rank considerations, since $\dim A = \dim V$, we have that it is also onto. Then we have the thesis by passing to bundles.

\end{proof} 
\end{lemma}

\begin{cor}
In the notations of the lemma above, any map $X \colon M \rarr VB$ has one and only one corresponding map $Y \colon M \rarr E$ with
\eq{
X(a_x) = [\gamma(a_x, Y(x))]_{a_x}
}
\end{cor}

Therefore, since vector bundles are also affine, we have
\eq{
\label{bi_dynamical_variations}
&V(T^*M \ot_M E) \simeq (T^*M \ot_M E) \times_M (T^*M \ot_M E)
	\\
&V(\Con_{\SU(2)}({}^+ Q)) \simeq (\Con_{\SU(2)}({}^+ Q)) \times_M (T^*M \ot_M ({}^+ Q \times_{\Ad} \su(2)) )
	\\
&V(T^*M \ot_M {}^+ E^\beta)) \simeq (T^*M \ot_M {}^+ E^\beta)) \times_M (T^*M \ot_M {}^+ E^\beta))
}

Denote by $e^{s X} \colon \sv C \rarr \sv C$ the $1$-parameter family of diffeomorphisms induced by the variation $X$, then for any first-order Lagrangian $\sv L \colon J^1 \sv C \rarr \Lambda^m M$ we define the \emph{variation of $\sv L$ along $X$} as $\delta_X \sv L$
\eq{
\delta_X \sv L 
&= \left. \der{}{s} (\sv L \circ j^1e^{s X} ) \right|_{s = 0}
}
where $j^1 e^{s X} \colon J^1 \sv C \rarr J^1 \sv C$ is the (first) jet prolongation of the bundle map $e^{sX} \colon \sv C \rarr \sv C$. Since we will be considering $\delta_X \sv L$ for any possible variation $X$, we will simply omit $X$ and write $\delta \sv L$.

We now use the following fact
\begin{lemma}
Consider $\Phi \in \Omega^{k,h}_H(Q, \ell)$, then we have
\eq{
\delta(\st \Phi) = \st (\delta \Phi)
}
\begin{proof}
%See \cref{app_var_calc_1}.
Recall that for
\eq{
\Phi = \frac{1}{h!} \Phi^{a_1 \dots a_h} \ot T_{a_1} \w \dots \w T_{a_h}, \quad \Phi^{a_1 \dots a_h} \in \Omega^k_H(Q)
}
we have
\eq{
(\st \Phi)^{a_{h+1} \dots a_m} = - \frac{1}{h!} \epsilon_{a_1}^\.{}_{\dots}^\.{}_{a_h}^\.{}^{a_{h+1} \dots a_m} \, \Phi^{a_1 \dots a_h}
}
and
\eq{
\st \Phi = \frac{1}{(m-h)!} (\st \Phi)^{a_{h+1} \dots a_m} \ot T_{a_{k+1}} \w \dots \w T_{a_m}
}
Since the Levi--Civita symbol is constant it satisfies $\delta \ep_{a_1 \dots a_m} = 0$. Then
\eq{
\delta(\st \Phi) 
&= \frac{1}{(m-h)!} \delta (\st \Phi)^{a_{h+1} \dots a_m} \ot T_{a_{k+1}} \w \dots \w T_{a_m}
	\\
&= -\frac{1}{(m-h)!} \frac{1}{h!} \delta \( \epsilon_{a_1}^\.{}_{\dots}^\.{}_{a_h}^\.{}^{a_{h+1} \dots a_m} \, \Phi^{a_1 \dots a_h} \) \ot T_{a_{k+1}} \w \dots \w T_{a_m}
	\\
&= -\frac{1}{(m-h)!} \frac{1}{h!} \epsilon_{a_1}^\.{}_{\dots}^\.{}_{a_h}^\.{}^{a_{h+1} \dots a_m} \, \delta \( \Phi^{a_1 \dots a_h} \) \ot T_{a_{k+1}} \w \dots \w T_{a_m}
	\\
&= \st (\delta \Phi)
}

\end{proof}
\end{lemma}

We now compute the variation of the BIH lagrangian, start with
\eq{
\delta \sv L_\gamma 
&= \frac{1}{4\bar G} \st \[ \delta F \ow \st_\gamma \theta^2 + \delta (\D A \kappa) \ow \st_\gamma \theta^2 \right. + 
	\\
&\quad + \left. \half 1 \delta[\kappa \w \kappa] \ow \st_\gamma \theta^2 + R \ow \st_\gamma \delta \theta^2 \]
}

To expand the formula above, we express the variations $\delta F, \delta (\D A \kappa), \delta [\kappa \w \kappa]$ and $\delta \theta^2$ in terms of the basic variations $\delta \theta, \delta A$, and $\delta \kappa$. Using that $\delta$ commutes with $d$
\eq{
\delta F 
&= \delta \( dA + \half 1 [A \w A] \)
	\\
&= d(\delta A) + \half 1 [\delta A \w A] + \half 1 [A \w \delta A] 	
	\\
&= d(\delta A) + [A \w \delta A] 	
}
As shown in \cref{bi_dynamical_variations}, the variation $\delta A$ is a \emph{tensorial} $1$-form of type $(\Ad, \su(2))$ so that we have
\eq{
\delta F = \D A (\delta A)
}
Similarly
\eq{
\delta (\D A \kappa) 
&= \delta \( dK + [A \w \kappa] \)
	\\
&= d(\delta \kappa) + [\delta A \w \kappa] + [A \w \delta \kappa] 	
}
Again from \cref{bi_dynamical_variations}, the variation $\delta \kappa$ is a tensorial $1$-form of type $(\Ad_{\Spin(3,1)}(\SU(2)), \lie m_\beta)$ so that we get
\eq{
\delta (\D A \kappa) = \D A (\delta \kappa) + [\delta A \w \kappa]
}
Then
\eq{
\delta [\kappa \w \kappa] 
&= 2 [\delta \kappa \w \kappa]
}
Finally
\eq{
\delta \theta^2
&= \delta (\theta \ow \theta)
	\\
&= 2 \theta \ow \delta \theta
}

By plugging the basic variations into the Lagrangian we arrive at
\eq{
\delta \sv L_\gamma 
&= \frac{1}{4\bar G} \st \[ \( \D A (\delta A) + \D A (\delta \kappa) + [\delta A \w \kappa] + [\delta \kappa \w \kappa] \) \ow \st_\gamma \theta^2+ 2(F + \D A \kappa) \ow \st_\gamma (\theta \ow \delta \theta) \]
	\\
&= \frac{1}{4\bar G} \st \[ \( \D A (\delta A) + \D A (\delta \kappa) + [\delta (A+\kappa) \w \kappa] \) \ow \st_\gamma \theta^2 + 2\st_\gamma R \ow \theta \ow \delta \theta \]
%}
%\red{not needed
%\begin{lemma}
%Consider $\Theta, \Phi \in \Omega^{1,2}(Q, \ell)$ and $\Psi \in \Omega^{2,2}(Q, \ell)$, then we have
%\eq{
%[\Theta \w \Phi] \ow \Psi = \Theta \dotw (\Phi \ow \Psi) + \Phi \ow (\Theta \dotw \Psi)
%}
%\end{lemma}
%
%\begin{proof}
%See \cref{app_var_calc_2}.
%\end{proof}
%}
%
%\red{not needed
%\begin{lemma}
%Consider $\Phi \in \Omega^{1,2}_H(Q, \ell)$ and $\Psi \in \Omega^{2,2}_H(Q, \ell)$, then we have
%\eq{
%\D A \Phi \ow \Psi = \D A(\Phi \ow \Psi) + \Phi \ow \D A \Psi
%}
%\end{lemma}
%}
%
%Using the lemmas we get
%\eq{
%\delta \sv L_\gamma 
	\\
&= \frac{1}{4\bar G} \st \[ \delta A \ow \D A \( \st_\gamma \theta^2 \) + \delta \kappa \ow \D A(\st_\gamma \theta^2) + 2\st_\gamma R \ow \theta \ow \delta \theta \right. + 	
	\\
&\quad + \left. [\delta (A+\kappa) \w \kappa] \ow \st_\gamma \theta^2  \right. + 
	\\
&\quad + \left. \D A \( \delta A \ow  \st_\gamma \theta^2  + \delta \kappa \ow  \st_\gamma \theta^2 \) \]
}

Since $\D A \st = \st \D A$ (\cref{vec_forms_hodge_prop}, item $(\rom 7)$) and and $\st_\gamma = (1 - \frac{1}{\gamma} \st)$, we get that also $\D A$ and $\st_\gamma$ commute. The last line in the expression above can be rewritten using
\begin{lemma}
Consider $\Theta, \Phi \in \Omega^{1, 2}(Q, \ell)$ and $\Psi \in \Omega^{2,2}(Q, \ell)$, then we have
\eq{
[\Theta \w \Phi] \ow \st \Psi = \Theta \ow \st[\Phi \w \Psi]
}
\begin{proof}

Let us assume, without loss of generality, that all elements are decomposable
\eq{
&\Theta = \tht \ot u
	\\
&\Phi = \pphi \ot v
	\\
&\Psi = \psi \ot w
}
then
\eq{
[\Theta \w \Phi] \ow \st \Psi
&= (\tht \w \pphi \w \psi) \ot \( [u, v] \w \st w \)
}
On $\Gamma(\Lambda^2 E)$ we have that $h$ is the Killing form for $[-, -]$ and $\st^2 = -\id$ so that
\eq{
[u, v] \w \st w
&= h([u, v], w) n_h
	\\
&= h(u, [v , w]) n_h
	\\
&= u \w \st [v, w] 
}
Therefore we get
\eq{
[\Theta \w \Phi] \ow \st \Psi 
&= (\tht \w \pphi \w \psi) \ot \( u \w \st[v, w] \)
	\\
&= \Theta \ow \st[\Phi \w \Psi]
}
which is the thesis.

\end{proof}
\end{lemma}

We can finally write
\eq{
\delta \sv L_\gamma 
&= \frac{1}{4\kappa} \st \[ (\delta A + \delta \kappa) \ow \st_\gamma (\D A \theta^2 + [\kappa \w \theta^2] ) + 2\st_\gamma R \ow \theta \ow \delta \theta \right. + 	
	\\
&\quad + \left. \D A \( \delta A \ow  \st_\gamma \theta^2  + \delta \kappa \ow  \st_\gamma \theta^2 \) \]
}

\subsection{E--L Equations for $\delta A$ and $\delta \kappa$}

We now use the Hamilton principle to get the Euler--Lagrange equations

\begin{defn}[Hamilton Principle (see \cite{fatifranca}, p.\ 154)]
Consider a configuration bundle $\sv C \rarr M$ on $M$, a first-order lagrangian $\sv L \colon J^1 \sv C \rarr \Lambda^m M$, and a variation $X \colon M \rarr V\sv C$, that is a section of the bundle of vertical vectors over $\sv C$ which is supported on a compact region $D \subset M$ and is zero on the boundary $\d D$. For any section $\sigma \colon M \rarr \sv C$, the \emph{action $\sv A_\sigma(D)$ of $\sv L$ on $D$} is the integral
\eq{
\sv A_\sigma(D) = \int_D L \circ j^1 \sigma
}
where $j^1 \sigma \colon M \rarr J^1 \sv C$ is the (first) jet prolongation of the section $\sigma$. The variation of $\sv A_\sigma(D)$ along $X$ is 
\eq{
\delta_X \sv A_\sigma(D) 
&= \int_D \delta_X \( \sv L \circ j^1 \sigma \) 
}

The Hamilton principle states that $\sigma \colon M \rarr \sv C$ is a \emph{critical section} (or \emph{classical solution}) if
\eq{
\delta_X \sv A_\sigma(D) = 0
}
for any compact region $D \subset M$ and any variation $X \colon D \rarr V\sv C$ supported on $D$. The equations satisfied by a critical section are called \emph{Euler--Lagrange equations for $\sv L$}.
\end{defn}

The variation of the BIH lagrangian is zero whenever the following system is satisfied:
\eq{
\sys{
\delta A \ow \st_\gamma (\D A \theta^2 + [\kappa \w \theta^2] ) = 0
	\\
\delta \kappa \ow \st_\gamma (\D A \theta^2 + [\kappa \w \theta^2] ) = 0
	\\
2 \st_\gamma R \ow \theta \ow \delta \theta = 0
}
}
or
\eq{
\sys{
-\bk{\delta A}{\bar \hod \st_\gamma (\D A \theta^2 + [\kappa \w \theta^2] } = 0
	\\
-\bk{\delta \kappa}{\bar \hod \st_\gamma (\D A \theta^2 + [\kappa \w \theta^2] } = 0
	\\
-\bk{\bar \hod (2\st_\gamma R \ow \theta)}{\delta \theta} = 0
}
}
%Since it is already known \red{cite} that the equations for $\delta \theta$ are equivalent to Palatini--Einstein field equations, l
Let us now focus on the first two equations. From Hamilton principle we need to consider an arbitrary variation $\delta A$, that is an arbitrary tensorial $1$-form valued valued in $\su(2) \subset \spin(3,1)$. Similarly we need to consider an arbitrary variation $\delta \kappa$, which is an arbitrary tensorial $1$-form valued in $\lie m_\beta \subset \spin(3,1)$. These facts imply that the form
\eq{
\bar \hod \st_\gamma (\D A \theta^2 + [\kappa \w \theta^2])
}
is valued in $\su(2)^\perp \cap (\lie m_\beta)^\perp$, where the orthogonal complements are with respect to the metric $q$. 

Since $q$ is definite both on $\su(2)$ and $\lie m_\beta$ and given that $\su(2) \cap \lie m_\beta = 0$ and $\su(2) \oplus \lie m_\beta = \spin(3,1)$, we must have
\eq{
&\su(2)^\perp \cap (\lie m_\beta)^\perp = 0
	\\
&\rimpl \bar \hod \st_\gamma (\D A \theta^2 + [\kappa \w \theta^2]) = 0
}
Using that $\bar \hod$ and $\st_\gamma$ are isomorphisms, we are left with
\eq{
\D A \theta^2 + [\kappa \w \theta^2] = 0
}
The two terms can be expanded further
\eq{
\D A (\theta \ow \theta) 
&= d \theta \ow \theta - \theta \ow d \theta + [A \w (\theta \ow \theta)]
	\\
&= d \theta \ow \theta + d\theta \ow \theta + (A \cw \theta) \ow \theta + \theta \ow (A \cw \theta)
	\\
&= 2 \, d \theta \ow \theta + 2 \, (A \cw \theta) \ow \theta 
	\\
&= 2 \D A \theta \ow \theta
}
and
\eq{
[\kappa \w \theta^2]
&= [\kappa \w (\theta \ow \theta)]
	\\
&= 2 (\kappa \cw \theta) \ow \theta
}
Since both $\D A \theta$ and $\kappa \cw \theta$ are in $\Omega^{2,1}_H(Q, \ell)$, we have that $k + h = 3 < 4$ so that from \cref{vec_forms_owtheta_inj} the map $- \ow \theta$ is injective. Therefore the equation reduces to 
\eq{
&\D A \theta^2 + [\kappa \w \theta^2] = 0
	\\
&\rimpl 2\( \D A \theta + \kappa \cw \theta \) \ow \theta = 0
	\\
&\rimpl \D A \theta + \kappa \cw \theta = 0
}
and, since $\D A \theta = \Theta_A = C_A \cw \theta$, we finally get
\eq{
C_A \cw \theta = - \kappa \cw \theta
}
Using \cref{spin_struc_levi_civita} and its corollary, we have that $\kappa = -C_A$, minus the contorsion tensor of $A$, which means that
\eq{
\omega = A + \kappa = A - C_A = \{ e \}
}
That is, the E--L equations for $\delta A$ and $\delta \kappa$ together imply that the spin connection $\omega$ which is solution of the equations is the Levi--Civita connection $\{ e \}$ of the spin frame $e$, which is yet to be determined by the remaining E--L equation.

Notice how this is true \emph{for any} value of the Immirzi parameter: whatever $\beta$ is choosen, the E--L equations will always by solved by $\kappa = -C_A$.

\subsection{E--L Equations for $\delta \theta$}

Starting from
\eq{
-\bk{\bar \hod (2\st_\gamma R \ow \theta)}{\delta \theta} = 0
}
We use Hamilton principle and consider an arbitrary variation $\delta \theta$, that is an arbitrary tensorial $1$-form valued in $\R^4$, therefore the equations are equivalent to
\eq{
&-\bar \hod (2\st_\gamma R \ow \theta) = 0
	\\
&\rimpl \gamma \, \bar \hod ( R \ow \theta) = \bar \hod (\st R \ow \theta)
}

Recall that for $v \in \Lambda^2 E_x$ and $x \in E_x$ we have
\eq{
\st(v \w x) 
&= \st v \chair x^{\flat_h}
}
But since $\st v \in \Lambda^2 E_x$, the contraction coincides with 
\eq{
\st v \chair x^{\flat_h} = \st v \. x
}
where the dot denotes the action of $\Lambda^2 E_x$ on $E_x$ derived from that of $\Lambda^2 \R^m \simeq \spin(r,s)$ on $\R^m$.

We then have, using the Bianchi identity $\D\omega \Theta_\omega = R \cw \theta$
\eq{
\st (\st R \ow \theta) = \st^2 R \cw \theta = -R \cw \theta = - \D \omega \Theta_\omega
}
The equation can be then rewritten as
\eq{
\gamma \, \bar \hod ( R \ow \theta) 
&= \st \bar \hod (\D \omega \Theta_\omega)
}
Using the E--L equations for $\delta A, \delta \kappa$ we know that $\Theta_\omega = 0$ therefore we are left with
\eq{
\bar \hod(R \ow \theta) = 0
}
Using the formula for the traces (\cref{vec_forms_trace}) we have that this equation is equivalent to
\eq{
\tr (\bar \hod R) = 0
}
Since $R, \bar \hod R \in \Omega^{2,2}_H(Q, \ell)$ and $2 + 2 = 4$ we have, from \cref{vec_forms_trace_inj}
\eq{
\tr(\bar \hod R ) = 0 \iff \theta \ow \bar \hod R = 0 \iff \tr R = 0
}
Since the trace of the Riemann tensor $R$ is the Ricci tensor $Ric$ we finally have the (vacuum) Einstein Field Equations
\eq{
Ric = 0
}
as we expected.

We have thus shown that the BIH lagrangian in a vacuum gives that, for any value of the Immirzi parameter $\beta$, the extrinsic spacetime field $\kappa$ must coincide with minus the contorsion $C_A$ of the Barbero--Immirzi connection $A$. This in turn implies that the spin connection $\omega = A + \kappa$ constructed from $A, \kappa$ has to be equal to th Levi--Civita connection $\{ e \}$ of the spin frame $e \colon Q \rarr L(M)$. The remaining equations then are equivalent to the Einstein field equation, as expected.

The geometrical clarity of the results is the consequence of having reformulated the BIH lagrangian using the language of $\Lambda \R^m$-valued forms and their calculus. This framework is evidently well-suited to treat the case of the BIH lagrangian coupled with boson or fermion (spinor) fields, from which we can already predict that the relation between $\kappa$ and $C_A$ will be modified by the presence of spinor terms.

\section{Conclusions and Perspectives}

In this work we gave a well-defined geometric framework for building BI connections on any given lorentzian spin manifold $M$ of dimension $m \geq 3$. The construction generalises what is found in the physics and mathematical physics literature and improves it by not requiring a fixed metric $g$ on $M$ and producing BI connections on the whole spacetime manifold $M$. The much discussed BI parameter $\beta$ comes out naturally and we proved that it is a feature {\it unique} to the $4$-dimensional case.

As stated in the introduction, the main motivation behind this work was to reformulate General Relativity: instead of using a metric $g$ and a spin connection $\omega$ as field variables, we rewrite the Lagrangian in terms of a spin frame $e$ and the pair $(A,K)$ for a fixed BI parameter $\beta$. Since this formulation is the classical starting point of LQG, it is worthwhile to study the variational Euler--Lagrange equations for $(e,A,K)$ and their relation to the Einstein field equations. The equations dependence on the BI parameter $\beta$ is also of interest, especially when we couple General Relativity with other fields (e.g.\ Klein--Gordon, Yang--Mills, Dirac).

Another application of this work is the investigation of the holonomy group of a BI connection $A$, which plays a central role in LQG. Samuel's argument in \cite{samuelbi} can be solved by looking at the relation between the holonomy group of $A$ and that of the original spin connection $\omega$, both on $M$ and on a spacelike submanifold $S$. One should also refer to the classification of metric lorentzian holonomies by Berger \cite{bergerHol} and Leistner \cite{leistnerHol} to better characterize the BI connections, this was suggested in private form.

It would also be interesting to study if and how this work relates to the other, non canonical, generalization of BI connection as proposed in \cite{bodendorfer}.

In \crefrange{vec_forms_vec_forms}{vec_forms_trace_sec} we developed a full-fledged calculus for differential forms on a spin bundle which are valued in $\Lambda \R^m$: this is made possible by the unique isomorphism between spin algebras $\spin(r,s)$ and the rank $(2,0)$ skew-symmetric tensors $\Lambda^2 \R^m$, where $m = r+s$. This material came to life as a companion to the main focus of the work, furnishing the geometrical tools needed to streamline the calculations in the variational analysis of the Holst Lagrangian and to better understand the geometric significance of the objects which appeared. We soon came to the realization that the geometrical content of this \q{appendix} was rich enough to warrant treatment of its own, especially the sections which show the delicate interplay between composite Hodge duals, Kulkarni--Nomizu products, and traces.

In \crefrange{bi_dynamical_holst}{bi_dynamical_bih_el} we perform the variation of the Barbero--Immirzi--Holst Lagrangian. Even though variational calculus is anything but new, one wishes to perform the calculations in the most straightforward manner possible and to gain insight from the resulting Euler--Lagrange equations, even in the very common situation where no explicit solution to the equations is available. The fields to be varied are the triple $(\theta, A, \kappa)$ of solder form $\theta$, Barbero--Immirzi connection $A$, and extrinsic spacetime field $\kappa$, recasting the Holst Lagrangian as dependent on this triple requires the global costruction of Barbero--Immirzi connections, so that the present work heavily builds on our recent article \cite{orizz_bi_conn}. The variational itself is a direct application of the calculus of vector-valued forms, and the simplicity of the resulting equations is in itself a demonstration of the power of this tool. The main result we prove is that there are two sets Euler--Lagrange equations. The first set relates the extrinsic spacetime field $\kappa$ to minus the contorsion $C_A$ of the Barbero--Immirzi connection, so that the spin connection $\omega = A + \kappa$ is always the Levi--Civita connection induced by the spin frame $e$/solder form $\theta$, what is surprising is that this result is independent on the kinematical Immirzi parameter $\beta$ and Holst parameter $\gamma$ in the Holst Lagrangian. The second set of equations is show to be equivalent to Einstein field equations, a fact which is expected but nonetheless important.

The results obtained from the variational analysis of the Barbero--Immirzi--Holst Lagrangian can be expanded in various directions as well. First, one can use the power and simplicity of the vector-valued forms formalism to find and study suitable analogues of the Holst Lagrangian in lorentzian manifolds of generic dimension $m > 4$. The main reason for this is two understand better if and why the case of dimension $m = 4$ is special, which is the only dimension in which one can have a non zero Immirzi parameter $\beta$. Second, it is of great interest to study the Holst Lagrangian coupled with matter, especially the Klein--Gordon Lagrangian for bosons and the Dirac Lagrangian for fermions/spinors. As we already pointed out, the $\beta$ and $\gamma$ parameters have no \emph{classical} effect on the Euler--Lagrange equations in the vacuum case (i.e.\ no coupling with matter), but it is expected that they play some role in matter couplings, particularly for spinors. This constitutes the direct continuation of this work, and one can also combine it with the first point and study matter couplings in a generic dimension $m$.

Finally, one can use the results of the \emph{classical} variational analysis as the starting point for quantization, this is what is done in Loop Quantum Gravity and is one of the principal reason we decided to investigate the Barbero--Immirzi connections and Holst Lagrangian in greater detail. Quantization can be performed either starting from the canonical analysis, or directly from fields on spacetime. The former approach leads to the spin network framework, while the latter leads to the spinfoam framework. In either case the quantization of a relativistic theory aims at defining suitable \q{discrete} analogues for geometric structures like manifolds, connections, and differential equations. Even though the discretization may seem an approximation, when one starts from smooth structures, the physical viewpoint is that it is the differentiable objects that have to be obtained as suitable limiting cases of the more fundamental, quantum, discrete objects. The correct, or rather most functional, definition of discrete geometry has not been worked out in full detail, and neither has the relationship between the discrete objects and their smooth counterparts. For this purpose, the work in this thesis constitutes one first, small step towards the understanding of discrete geometric structure or, in physics parlance, quatum gravity.

%\begin{acknowledgements}
\section*{Acknowledgements}

The author acknowledges the contribution of INFN (Iniziativa Specifica QGSKY), the local research project {\it  Metodi Geometrici in Fisica Matematica e Applicazioni (2019)} of Dipartimento di Matematica of University of Torino (Italy). This paper is also supported by INdAM-GNFM.

I would also like to thank Lorenzo Fatibene for its precious insights and suggestions, which greatly improved the exposition of the material covered here.

%\renewcommand{\appendixpagename}{Appendices}
%\renewcommand{\appendixtocname}{Appendices}
%\appendixpage
\appendix
\section{Trace Lemma}\label[app]{app_trace_lemma}

We now prove two results which were stated in \cref{vec_forms_trace_sec}. The first is

\begin{lemma}[Trace Lemma]
For any $\Phi \in \Omega^{k,h}_H(Q, \ell)$ we have the following identity
\eq{
\tr (\theta \ow \Phi) = \theta \ow \tr \Phi  + (m - k - h) \Phi
}
\begin{proof}
We prove this for $\Phi$ totally decomposable, that is
\eq{
\Phi = (\pphi_1 \w \dots \w \pphi_k) \ot (v_1 \w \dots \w v_h)
}
with $\pphi_i \in \Omega^1(M)$ and $v_j \in \Gamma(E)$. Since $\tr \Phi = \Phi \chair \theta^\sharp$ we first consider $\{ T_a \}$, the $\eta$-orthonormal basis of $\R^m$, and its dual basis $\{ \tau^a \}$. Then we can decompose
\eq{
\theta &= \theta^a \ot T_a, \quad \theta^a \in \Omega^1(M)
	\\
\theta^\sharp &= e_a \ot \tau^a, \quad e^a \in \lie X(M)
}
Therefore
\eq{
\tr \Phi = \Phi \chair \theta^\sharp
&= (e_a \ot \tau^a) \chair [(\pphi_1 \w \dots \w \pphi_k) \ot (v_1 \w \dots \w v_h)]
	\\
&= [e_a \chair (\pphi_1 \w \dots \w \pphi_k)] \ot [\tau^a \chair (v_1 \w \dots \w v_h)]	
	\\
&= \left. \sum_{i = 1}^k (-1)^{i -1} (e_a \chair \pphi_i) \, \pphi_1 \w \dots \w \check \pphi_i \w \dots \w \pphi_k \right. \ot 
	\\
&\quad \ot \left. \sum_{j = 1}^h (-1)^{j-1}(\tau^a \chair v_j) \, v_1 \w \dots \w \check v_j \w \dots \w v_h \right.
}

The form $\theta \ow \Phi$ is
\eq{
\theta \ow \Phi = (\theta^b \w \pphi_1 \w \dots \w \pphi_k) \ot (T_b \w v_1 \w \dots \w v_h)
}
Denote by $\pphi_0 = \theta^b$ and $v_0 = T_b$, then
\eq{
\tr(\theta \ow \Phi)
&= \left. \sum_{i = 0}^k (-1)^{i} (e_a \chair \pphi_i) \, \pphi_0 \w \dots \w \check \pphi_i \w \dots \w \pphi_k \right. \ot 
	\\
&\quad \ot \left. \sum_{j = 0}^h (-1)^{j}(\tau^a \chair v_j) \, v_0 \w \dots \w \check v_j \w \dots \w v_h \right.
}
The term with $i = j = 0$ is
\eq{
&[(e_a \chair \theta^b) \, \pphi ] \ot [(\tau^a \chair T_b) \, v]
	\\
&= (\theta \chair \theta^\sharp) \Phi
	\\
&= m \, \Phi
}
where we used $\theta \chair \theta^\sharp = \bk{\theta}{\theta} = m$.

There are $h$ terms with $i = 0$ and $j \neq 0$, which are
\eq{
&(\delta_a^b \, \pphi) \ot [(-1)^{j}(\tau^a \chair v_j) \, T_b \w v_1 \w \dots \w \check v_j \w \dots \w v_h]
	\\
&= \pphi \ot [(-1)^{j}(\tau^a \chair v_j) \, T_a \w v_1 \w \dots \w \check v_j \w \dots \w v_h]
	\\
&= \pphi \ot [(-1)^{j} \, v_j \w v_1 \w \dots \w \check v_j \w \dots \w v_h]
	\\
&= \pphi \ot (-v_1 \dots \w v_j \w \dots \w v_h)
	\\
&= - \Phi	
}
so that they sum to $-h \Phi$.

There are $k$ terms with $i \neq 0$ and $j = 0$, which are
\eq{
&[(-1)^{i} (e_a \chair \pphi_i) \, \theta^b \w \pphi_1 \w \dots \w \check \pphi_i \w \dots \w \pphi_k] \ot (\delta^a_b \, v)
	\\
&= [(-1)^{i} (e_a \chair \pphi_i) \, \theta^a \w \pphi_1 \w \dots \w \check \pphi_i \w \dots \w \pphi_k] \ot v
	\\
&= [(-1)^{i} \pphi_i \w \pphi_1 \w \dots \w \check \pphi_i \w \dots \w \pphi_k] \ot v	
	\\
&= (-\pphi_1 \w \dots \w \pphi_i \w \dots \w \pphi_k) \ot v		
	\\
&= -\Phi
}
so that they sum to $-k \Phi$.

Finally there are all the terms with $i \neq 0 \neq j$
\eq{
& \[ \theta^b \w \sum_{i = 1}^k (-1)^{i} (e_a \chair \pphi_i) \, \pphi_1 \w \dots \w \check \pphi_i \w \dots \w \pphi_k \] \ot 
	\\
&\qquad \ot \[ T_b \w \sum_{j = 1}^h (-1)^{j}(\tau^a \chair v_j) \, v_1 \w \dots \w \check v_j \w \dots \w v_h \]
	\\
&= (-1)\[ \theta^b \w \sum_{i = 1}^k (-1)^{i-1} (e_a \chair \pphi_i) \, \pphi_1 \w \dots \w \check \pphi_i \w \dots \w \pphi_k \] \ot 
	\\
&\qquad \ot (-1)\[ T_b \w \sum_{j = 1}^h (-1)^{j-1}(\tau^a \chair v_j) \, v_1 \w \dots \w \check v_j \w \dots \w v_h \]	
	\\
&= \theta \ow \tr \Phi
}

By adding all the terms, we get the thesis.

\end{proof}
\end{lemma}

The second result is

\begin{prop}[Injectivity of the Trace]
\label{vec_forms_trace_inj}
For $k + h > m$ the map
\eq{
\map{\tr}{\Omega^{k,h}_H(Q, \ell)}{\Omega^{k-1,h-1}_H(Q, \ell)}{\Phi}{\tr \Phi}
}
is injective. In particular, for $q = \min\{ m-k, m-h \}$ we have
\eq{
\tr \Phi = 0 \rimpl \theta^i \ow \Phi = 0, \quad \forall i = 0, \dots, q
}
while for $k + h = m \iff p = - m + k + h = 0$ we have
\eq{
\tr \Phi = 0 \rimpl \theta^i \ow \Phi = 0, \quad \forall i = 1, \dots, q
}
\begin{proof}
We compute the iterated powers of $\theta^r \ow \Phi$ by induction:
\eq{
\theta^{r+1} \ow \tr \Phi = -(r+1)(m - k - h - r) \theta^r \ow \Phi + \tr(\theta^{r+1} \ow \Phi)
}
The base case $r = 0$ is the previous lemma, then
\eq{
\theta^{r+1} \ow \Phi &= \theta \ow (\theta^r \ow \Phi)
	\\
&= \theta \ow (-r(m - k - h - r + 1) \theta^{r-1} \ow \Phi) + \theta \ow \tr(\theta^r \ow \Phi)
	\\
&= -r(m - k - h - r + 1) \theta^r \ow \Phi + \tr(\theta^{r+1} \ow \Phi) - (m - (k+r) - (h+r)) \theta^r \ow \Phi
	\\
&= -[(r+1)(m - k - h) - r(r-1) - 2r] \theta^r \ow \Phi + \tr(\theta^r \ow \Phi)
	\\
&= -[(r+1)(m - k - h) - r(r+1)] \theta^r \ow \Phi + \tr(\theta^r \ow \Phi)
	\\
&= -(r+1)(m - k - h - r) \theta^r \ow \Phi + \tr(\theta^r \ow \Phi)
}
which concludes the induction.

If $\tr \Phi = 0$, $-m + k + h = p$ and $q = \min\{ m - k, m - h\}$ then the identities above reduce to
\eq{
-(r+1)(p + r) \theta^r \ow \Phi = \tr(\theta^{r+1} \ow \Phi)
}
which gives
\eq{
&-p\Phi = \tr(\theta \ow \Phi)
	\\
&-2(p + 1) \theta \ow \Phi = \tr(\theta^2 \ow \Phi)	
	\\
& \qquad \vdots
	\\
&-q(p + q - 1) \theta^{q-1} \ow \Phi = \tr(\theta^{q} \ow \Phi)
	\\
&-(q+1)(p + q) \theta^q \ow \Phi = 0
}
If $p + q = 0$ then either $k = 0$ or $h = 0$, which implies that $k > m$ or $h > m$ (since $k + h > m$) therefore $\theta \ow \Phi$ is always zero and there is nothing to prove. Otherwise $p + q > 0$ (we are using that $k + h > m \iff p > 0$) and we can solve the system of equation backwards
\eq{
\Phi = \theta \ow \Phi = \dots = \theta^q \ow \Phi = 0
}
which is the thesis.

For $k + h = m \iff p = 0$ we get the partial result
\eq{
\tr \Phi = 0 \rimpl \theta^i \ow \Phi = 0, \quad i = 1, \dots, q
}

\end{proof}
\end{prop}

\newpage
\bibliographystyle{alpha}
%\bibliography{C:/Users/andre/Desktop/university/phd/tex/bib} 
\bibliography{bib} 

\end{document}